\documentclass[preprint,12pt]{elsarticle}

\usepackage{graphicx} 
\usepackage{enumerate}
\usepackage{subfigure}
\usepackage{tabularx}
\usepackage{caption}
\usepackage{hyperref}
\usepackage{float}
\usepackage{tabularx}
\usepackage{makecell}
\usepackage{multirow}
\usepackage{booktabs}
\usepackage{amsfonts} 
\usepackage{graphicx}
\usepackage{amsmath}
\usepackage{color}
\usepackage{caption}
\usepackage[misc]{ifsym} 
\usepackage{amsmath,amssymb,amsfonts}
\usepackage{algorithmic}
\usepackage{graphicx}
\usepackage{textcomp}

\journal{Medical Image Analysis}

\begin{document}

\begin{frontmatter}



\title{VHU-Net: Variational Hadamard U-Net for Body MRI Bias Field Correction}


\author[inst1,inst2]{Xin Zhu\texorpdfstring{\corref{1}}{textcorref}}
\ead{xzhu61@uic.edu}
\cortext[1]{Corresponding author}
\author[inst2]{Ahmet Enis Cetin}
\author[inst1]{Gorkem Durak}
\author[inst3]{Batuhan Gundogdu}
\author[inst1]{Ziliang Hong}
\author[inst1]{Hongyi Pan}
\author[inst1]{Ertugrul Aktas}
\author[inst1]{Elif Keles}
\author[inst1]{Hatice Savas}
\author[inst3]{Aytekin Oto}
\author[inst1]{Hiten Patel}
\author[inst1]{Adam B. Murphy}
\author[inst1]{Ashley Ross}
\author[inst1]{Frank Miller}
\author[inst4]{Baris Turkbey}
\author[inst1]{Ulas Bagci}

\affiliation[inst1]{organization={Machine and Hybrid Imaging Lab},
    addressline={Northwestern University}, 
    city={Chicago},
    country={USA}}

\affiliation[inst2]{organization={Department of Electrical and Computer Engineering},
    addressline={University of Illinois Chicago}, 
    city={Chicago},
    country={USA}}

\affiliation[inst3]{organization={Department of Radiology},
    addressline={University of
Chicago}, 
    city={Chicago},
    country={USA}}

\affiliation[inst4]{organization={Molecular Imaging Branch, NCI},
    addressline={National Institutes of Health}, 
    city={Bethesda, MD},
    country={USA}}

\begin{abstract}
Bias field artifacts in magnetic resonance imaging (MRI) scans introduce spatially smooth intensity inhomogeneities that degrade image quality and hinder downstream analysis. To address this challenge, we propose a novel variational Hadamard U-Net (VHU-Net) for effective body MRI bias field correction. The encoder comprises multiple convolutional Hadamard transform blocks (ConvHTBlocks), each integrating convolutional layers with a Hadamard transform (HT) layer. Specifically, the HT layer performs channel-wise frequency decomposition to isolate low-frequency components, while a subsequent scaling layer and semi-soft thresholding mechanism suppress redundant high-frequency noise. To compensate for the HT layer's inability to model inter-channel dependencies, the decoder incorporates an inverse HT-reconstructed transformer block, enabling global, frequency-aware attention for the recovery of spatially consistent bias fields. The stacked decoder ConvHTBlocks further enhance the capacity to reconstruct the underlying ground-truth bias field. Building on the principles of variational inference, we formulate a new evidence lower bound (ELBO) as the training objective, promoting sparsity in the latent space while ensuring accurate bias field estimation. Comprehensive experiments on body MRI datasets demonstrate the superiority of VHU-Net over existing state-of-the-art methods in terms of intensity uniformity. Moreover, the corrected images yield substantial downstream improvements in segmentation accuracy. Our framework offers computational efficiency, interpretability, and robust performance across multi-center datasets, making it suitable for clinical deployment. The codes are available at \url{https://github.com/Holmes696/Probabilistic-Hadamard-U-Net}.
\end{abstract}







\begin{keyword}
Bias field correction, Hadamard transform, U-Net, variational inference, MRI segmentation
\end{keyword}

\end{frontmatter}



\section{Introduction}
\label{sec:introduction}
\begin{figure}[htbp]
	\centering
		\includegraphics[scale=.39]{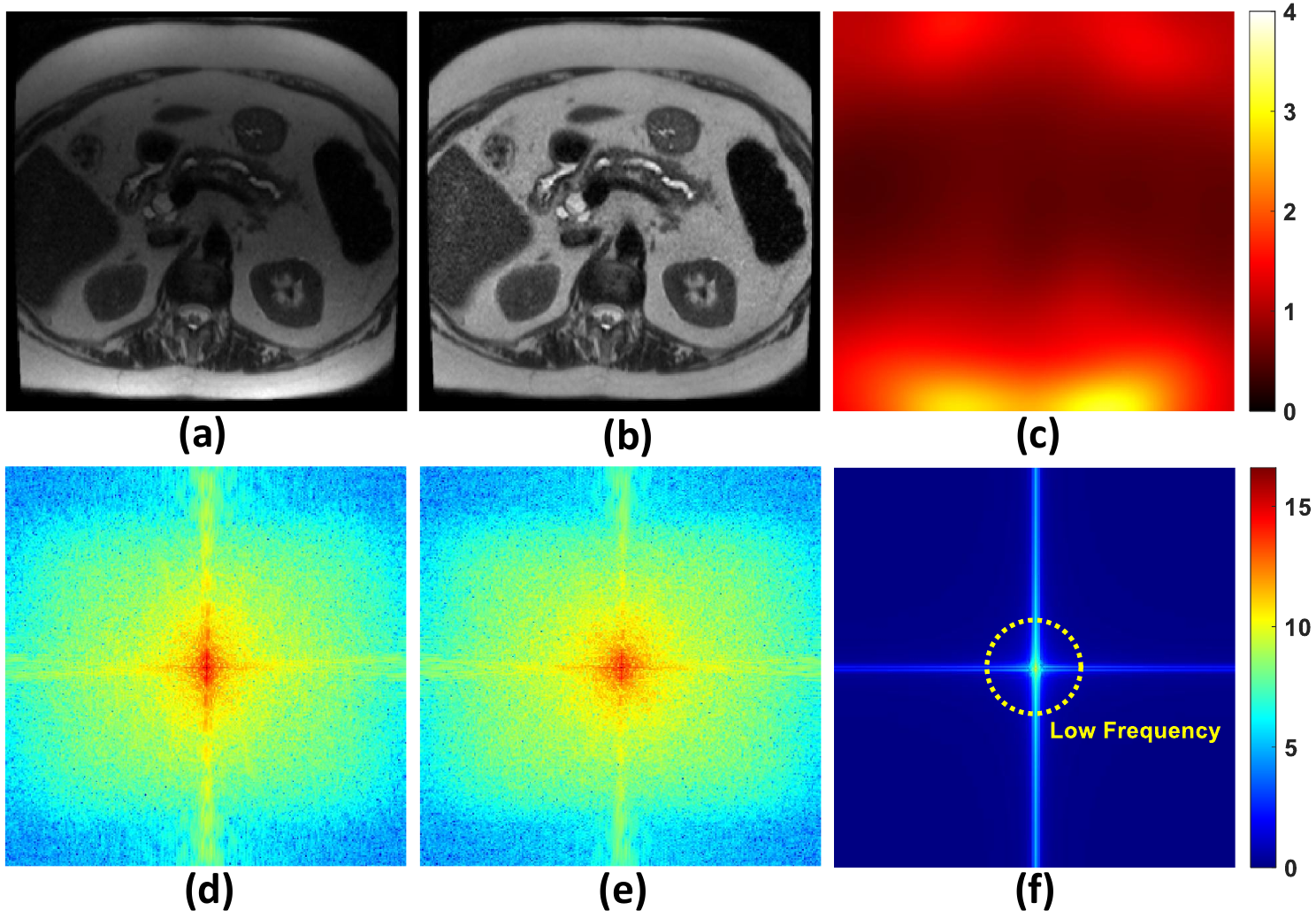}
	\caption{(a) Observed abdominal MRI image; (b) Bias-corrected image; (c) Bias field; (d–f) Frequency spectra corresponding to (a–c), respectively.}
	\label{Fig:timefrequency}
\end{figure}

The bias field represents a smooth, low-frequency multiplicative distortion arising from inhomogeneities in the primary magnetic field of magnetic resonance imaging (MRI) systems~\cite{moghadasi2021segmentation,jagadeesh2024brain}.  As illustrated in Fig.~\ref{Fig:timefrequency} (a), this distortion alters the intensity values in MRI images and produces heterogeneous gray-level distributions for the same tissue. 
Such variability in tissue intensity obscures fine details critical for accurate clinical interpretation. Consequently, the bias field poses challenges for a range of quantitative analyses, including efficient detection, accurate diagnosis, and robust segmentation~\cite{kanakaraj2024deepn4}.

To address this challenge, various sophisticated methods have been developed for bias field correction. Contemporary bias field correction methodologies can be classified into two primary categories: traditional methods and neural network-based deep learning methods.
Among the most well-known traditional methods, N4ITK~\cite{tustison2010n4itk} builds upon the original N3~\cite{sled2002nonparametric} algorithm by integrating a robust B-spline approximation and an advanced hierarchical optimization framework. This methodology has demonstrated superior efficacy in bias field correction for brain images, predicated on the assumption of intensity uniformity across distinct brain tissues~\cite{chen2021abcnet}. 
However, this assumption does not hold in body MRI, particularly in abdominal and prostate imaging. These regions contain complex anatomical structures and heterogeneous tissues, leading to substantial intensity variations as shown in Fig.~\ref{Fig:timefrequency} (b). 
As a result, the performance of N4ITK is compromised in these scenarios. Moreover, N4ITK removes the bias field at the expense of significant computational costs, resulting in a low inference speed~\cite{zhu2024probabilistic}.

In recent years, neural network-based deep learning methods have been widely used in addressing bias field correction. 
Smiko~\textit{et al.} introduced an implicit training method for MRI bias field removal, bypassing the need for ground-truth data and enabling non-medical data for model training~\cite{simko2022mri}. This approach generalized better across various anatomies and significantly outperformed traditional methods such as N4ITK in both speed and performance. 
Besides, Sridhara~\textit{et al.} proposed a deep learning approach leveraging an autoencoding framework to estimate bias fields in MRI images
~\cite{sridhara2021bias}. By correcting intensity inhomogeneities, this method improved tissue classification accuracy, outperforming traditional histogram-based methods when evaluated against ground-truth results. In addition, Chen~\textit{et al.} introduced a novel end-to-end 3D adversarial bias correction network (ABCnet) for infant brain MRI analysis, addressing challenges posed by the dynamic, heterogeneous intensity contrasts caused by nonuniform myelination~\cite{chen2021abcnet}. 
The ABCnet was trained alternately to minimize generative and adversarial losses, incorporating a tissue-aware local intensity uniformity term for reducing intensity variation, alongside terms for bias field smoothness and method robustness. It improved accuracy and efficiency in both simulated and real infant brain datasets. 
However, these approaches have notable limitations: (1) They often require multiple types of annotations, such as bias field maps and tissue segmentations, which are difficult to obtain in body MRI. (2) They typically rely on leveraging the globally smooth and spatially coherent nature of bias fields to facilitate estimation. However, the inherent complexity of body MRI disrupts such global trends. Anatomical regions such as the abdomen and prostate exhibit diverse and heterogeneous tissue properties, leading to pronounced inter- and intra-organ intensity variations that complicate the separation of tissue structures and bias fields.
Moreover, the MeMGB-Diff~\cite{qiu2025memgb} adopts a self-supervised learning strategy by modeling the three-dimensional bias field as a multivariate Gaussian and randomly generating synthetic bias fields during training. This approach effectively approximates the statistical properties of real bias artifacts in brain MRI specifically. Nevertheless, its generalizability to body MRI remains limited. The simulated bias fields used for training cannot adequately capture the complex and heterogeneous bias distributions present in real body MRI data, which restricts the robustness and applicability of the model beyond brain imaging.
Hence, comprehensive studies on neural network-based methods for body MRI bias correction are urgently needed but still lacking.


The extraction of the low-frequency multiplicative field is a fundamental step in MRI bias field correction, as it substantially influences the quality of the corrected images. Effective frequency-domain analysis methods are essential for optimizing this process. Currently, prominent approaches for frequency-domain analysis encompass the discrete Fourier transform~\cite{chi2020fast}, discrete cosine transform~\cite{zhu2025edge}, discrete wavelet transform~\cite{zhu2025efficient}, and Hadamard transform (HT)~\cite{zhu2025unsupervised}. These transformations are extensively applied to isolate distinct frequency-domain features, particularly low-frequency components critical for bias field correction. Among these methods, the HT stands out for its computational efficiency, as its transformation matrix consists solely of elements valued at $+1$ or $-1$, significantly simplifying implementation~\cite{fan2025real}.  
Despite the demonstrated effectiveness of orthogonal transforms in feature extraction tasks, their integration with advanced architectures, such as U-Nets, for MRI bias field correction remains unexplored.

To address the limitations of existing bias field correction methods, this paper proposes a novel frequency-aware variational Hadamard U-Net (VHU-Net) for body MRI bias correction. Conventional deep learning approaches infer the bias field by exploiting its global spatial smoothness. In contrast, our method projects MRI data into the Hadamard domain to directly capture the bias field more effectively.
Moreover, a single convolutional layer cannot be efficiently trained in tandem with a semi-soft thresholding nonlinearity. By placing a fixed HT after the convolutional layer, VHU-Net enables joint
training of both the convolutional filters and the threshold parameters, which adapts the data and enhances the capacity of the HT to isolate low-frequency bias fields.
Specifically, the encoder module of VHU-Net processes each input through a series of convolutional Hadamard transform blocks (ConvHTBlocks) to obtain an effective latent representation.
Subsequently, the decoder recovers the bias field through an inverse HT-reconstructed transformer block (IHTRTB) and a symmetric set of ConvHTBlocks.
In addition, a lightweight hypernetwork modulates the decoder to handle distributional shifts and improve adaptation under varying correction scenarios. The entire correction process is regularized via an evidence lower bound (ELBO)-driven loss function that ensures stable and domain-robust representation learning.
We extend our prior Probabilistic Hadamard U-Net (PHU-Net) framework (published in MICCAI MLMI 2024~\cite{zhu2024probabilistic}), with several important methodological enhancements:
\begin{enumerate}[1)]
\item We introduce a novel ConvHTBlock that combines spatial domain convolutions with a HT layer, which isolates frequency components in a channel-wise manner. Each HT layer integrates a trainable scaling layer and a semi-soft thresholding operator to selectively preserve low-frequency bias-related components while attenuating high-frequency redundancies, thereby facilitating salient feature extraction.
\item To address the HT layer’s deficiency in modeling inter-channel dependencies, we introduce an IHTRTB in the decoder. This module applies a transformer layer to capture global frequency-aware attention patterns and incorporates a semi-soft thresholding layer to further refine the denoising process. This combination enables accurate and spatially coherent bias field reconstruction.
\item A new probabilistic framework based on variational inference is developed, wherein the ELBO acts as the core optimization objective. This formulation enforces sparsity in the latent Hadamard space by constraining the approximate posterior to follow a Laplacian distribution with a small scale parameter, leading to improved robustness and generalizability of the model.
\item We conduct comprehensive evaluations on body MRI datasets encompassing abdominal, prostate, and breast regions. The proposed VHU-Net consistently surpasses state-of-the-art bias correction methods, delivering superior intensity homogeneity and notable improvements in downstream segmentation performance. In addition, the model exhibits fast inference speed, ensuring practical applicability in clinical workflows.
\end{enumerate}  


\section{Related Work}

\subsection{Bias Field Modeling}
A variety of techniques have been developed to tackle intensity nonuniformity in MRI images, with the problem typically  formulated as~\cite{perez2024unsupervised,qiu2025memgb}:
\begin{equation}
\mathbf{r} =  \mathbf{b} \circ \mathbf{f} + \mathbf{n},
\end{equation}
where $\mathbf{r}$ denotes the observed MRI image, $\mathbf{b}$ is the multiplicative bias field, $\mathbf{f}$ is the underlying inhomogeneity-free image, and $\mathbf{n}$ represents additive noise. The operator $\circ$ denotes the element-wise product.
In contemporary machine learning paradigms, the bias field is commonly modeled as the output \(\mathbf{b}_\alpha \) of a neural network parameterized by learnable weights \( \alpha \). The correction task is formulated as a supervised learning problem, where the loss function is defined as:
\begin{equation}
\mathcal{L}(\alpha) = \mathcal{L}_{\text{recon}}(\mathbf{r}, \mathbf{b}_\alpha \circ \mathbf{f}) + \lambda \mathcal{R}(\mathbf{b}_\alpha),
\label{eq:LR}
\end{equation}
where \( \mathcal{L}_{\text{recon}} \) represents reconstruction error with respect to the observations. \( \mathcal{R} \) is a smoothness regularization term. \( \lambda \) is the regularization weight.
However, obtaining inhomogeneity-free MRI images remains a challenge as high-quality phantom data is extremely scarce. To address this, many established methods~\cite{chen2021abcnet,zhu2024probabilistic} utilize the well-regarded bias correction method (e.g., N4ITK) as the reference standard. It provides ``cleaned" MRIs that serve as a substitute for ground truth. The learning-based formulation enables the model to adaptively capture complex variations in the bias field that may not be well approximated by classical parametric approaches, thus enhancing robustness across diverse acquisition protocols and anatomical regions.


\subsection{The Hadamard Transform}
The two-dimensional Hadamard transform (HT2D) is a discrete linear transform widely used in signal and image processing due to its computational efficiency and energy compaction properties~\cite{xia2025adaptive,thi2024efficient}. It is based on the Hadamard matrix, $\mathbf{H}_N$, an orthogonal matrix of size $N\times N$, which is recursively constructed as follows:
\begin{equation}
\mathbf{H}_1=1, 
\mathbf{H}_2 = \begin{bmatrix}
1 & 1 \\
1 & -1
\end{bmatrix}, 
\mathbf{H}_N = \begin{bmatrix}
\mathbf{H}_{N/2} & \mathbf{H}_{N/2} \\
\mathbf{H}_{N/2} & -\mathbf{H}_{N/2}
\end{bmatrix},
\end{equation}
where $N=2^k$, $k\in \mathbb{N}$. $\mathbf{H}_1$ is the base case, and $\mathbf{H}_N$ is constructed iteratively from $\mathbf{H}_{N/2}$. For a given input $\mathbf{x} \in \mathbb{R}^{N \times N}$, the HT2D is computed by applying the HT to both rows and columns of $\mathbf{x}$. The transformation is expressed as:
\begin{equation}
\label{Eq: HT}
\mathbf{Y} = \mathbf{H}_N \mathbf{x} \mathbf{H}_N,
\end{equation}
where $\mathbf{Y} \in \mathbb{R}^{N \times N}$ denotes the HT coefficients of $\mathbf{x}$. A notable property of HT is its capacity to concentrate most of the signal's energy into a small subset of coefficients, primarily located in the upper-left region of $\mathbf{Y}$~\cite{chen2022blind}. This property makes the HT2D effective for low-frequency feature extraction. To reconstruct original matrix $\mathbf{x}$, the inverse HT2D is defined as:
\begin{equation}
\label{Eq: IHT}
\widetilde{\mathbf{x}} = \frac{1}{N^2} \mathbf{H}_N \mathbf{Y} \mathbf{H}_N,
\end{equation}
where $\widetilde{\mathbf{x}} \in \mathbb{R}^{N \times N}$ represents the reconstructed output. In this study, we utilize the HT2D to isolate the low-frequency components of the body MRI image, which capture the bias field characteristics as shown in Fig.~\ref{Fig:timefrequency} (f).

\subsection{Variational Inference}
Variational inference has emerged as a powerful framework for approximate Bayesian inference, particularly in settings where exact posterior computation is intractable~\cite{cai2024eigenvi}. Specifically, variational inference introduces a family of tractable distributions and seeks the member that minimizes the Kullback–Leibler (KL) divergence to the true posterior~\cite{lim2024particle}. This formulation enables scalable and efficient uncertainty-aware inference, making it compatible with deep learning models in medical imaging. 
For instance, integrating variational inference into neural network architectures allows for learning distributions over latent representations, thereby facilitating tasks such as probabilistic segmentation~\cite{kohl2018probabilistic}, uncertainty quantification~\cite{hosseinzadeh2023uncertainty}, and image synthesis~\cite{friedrich2024deep}. However, despite its demonstrated success in segmentation and related domains~\cite{bhat2022generalized}, the potential of variational inference remains underexplored in broader computational medical imaging tasks, such as MRI bias field correction.

\section{Methodology}

\subsection{Problem Formation}
\label{sec:formation}
To enable efficient body MRI bias field correction, we adopt a probabilistic formulation based on latent variable modeling. Let $\mathbf{x}$ denote the observed input image corrupted by intensity inhomogeneity, and ${\mathbf{z}}$ represent the corresponding ground-truth bias field. To learn a deep learning model that most plausibly generates ${\mathbf{z}}$ conditioned on $\mathbf{x}$, we aim to maximize the marginal log-likelihood $\log p({\mathbf{z}} | \mathbf{x})$ during the training stage, given by:
\begin{equation}
    \log p(\mathbf{z} | \mathbf{x}) = \log \int p(\mathbf{z}, \mathbf{y} | \mathbf{x}) \, d\mathbf{y},
\label{eq:pzyx}
\end{equation}
where $\mathbf{y}$ denotes a latent variable capturing the underlying structure of the image domain. However, this marginalization is generally intractable due
to the complexity of the joint distribution $p(\mathbf{z}, \mathbf{y} | \mathbf{x})$. To address this, 
we employ the variational inference framework by introducing an approximate posterior $q(\mathbf{y}|\mathbf{x}) $. This distribution is modeled by the encoder network, which infers the latent code $\mathbf{y}$ from the bias-corrupted input $\mathbf{x}$. Next, the marginal log-likelihood is rewritten as follows:
\begin{equation}
    \log p(\mathbf{z} | \mathbf{x}) = \log \int q(\mathbf{y} | \mathbf{x}) \cdot \frac{p(\mathbf{z}, \mathbf{y} | \mathbf{x})}{q(\mathbf{y} | \mathbf{x})} \, d\mathbf{y}.
\end{equation}

Applying Jensen’s inequality to the logarithm yields a lower bound $\mathcal{E}$ on the marginal log-likelihood:
\begin{equation}
    \log p(\mathbf{z} | \mathbf{x}) \geq \mathbb{E}_{q(\mathbf{y} | \mathbf{x})} \left[ \log \frac{p(\mathbf{z}, \mathbf{y} | \mathbf{x})}{q(\mathbf{y} | \mathbf{x})} \right]=\mathcal{E}.
        \label{eq:Jensen}
\end{equation}

We further decompose the joint distribution $p({\mathbf{z}}, \mathbf{y} | \mathbf{x})$ using the chain rule:
\begin{equation}
    p(\mathbf{z}, \mathbf{y} | \mathbf{x}) = p(\mathbf{z} | \mathbf{y}, \mathbf{x}) \cdot p(\mathbf{y} | \mathbf{x}),
    \label{eq:chain}
\end{equation}
where $p(\mathbf{y} | \mathbf{x})$ is the true posterior. Substituting Eq.~(\ref{eq:chain}) into Eq.~(\ref{eq:Jensen}), we obtain the evidence lower bound (ELBO):
\begin{align}
    \mathcal{E}&= \mathbb{E}_{q(\mathbf{y} | \mathbf{x})} \left[ \log p(\mathbf{z} | \mathbf{y}, \mathbf{x}) \right] 
    + \mathbb{E}_{q(\mathbf{y} \mid \mathbf{x})} \left[\log \frac{p(\mathbf{y} | \mathbf{x})}{q(\mathbf{y} | \mathbf{x})} \right] \nonumber \\
    &=\mathbb{E}_{q(\mathbf{y} | \mathbf{x})} \left[ \log p(\mathbf{z} | \mathbf{y}, \mathbf{x}) \right] 
    -\mathcal{D}_{KL}\left( q(\mathbf{y} | \mathbf{x}) \,\|\, p(\mathbf{y} | \mathbf{x}) \right),
\label{eq:elbo}
\end{align}
where the distribution $p(\mathbf{z} | \mathbf{y}, \mathbf{x})$ describes the probability that the decoder reconstructs the ground-truth bias field $\mathbf{z}$ from $\mathbf{y}$ and $\mathbf{x}$. The expected log-likelihood of this distribution quantifies the bias field estimation accuracy. 
Moreover, the second term represents the negative KL divergence, 
which measures the discrepancy between the approximate posterior $q(\mathbf{y} |\mathbf{x})$ and the true posterior $p(\mathbf{y} | \mathbf{x})$. 
 
Unlike the intractable marginal likelihood in Eq.~(\ref{eq:pzyx}), the ELBO is computationally feasible and provides a practical surrogate objective for model optimization. A detailed computation of the ELBO is provided in Section~\ref{sec:elboc}.
According to Eq.~(\ref{eq:elbo}), the key to effective bias field estimation lies in maximizing the ELBO. It can be achieved by promoting accurate reconstruction of ground-truth bias field while simultaneously minimizing the divergence between the approximate and true posteriors by enforcing structured regularization on the latent representations.

\begin{figure*}[t]
	\centering
		\includegraphics[scale=.21]{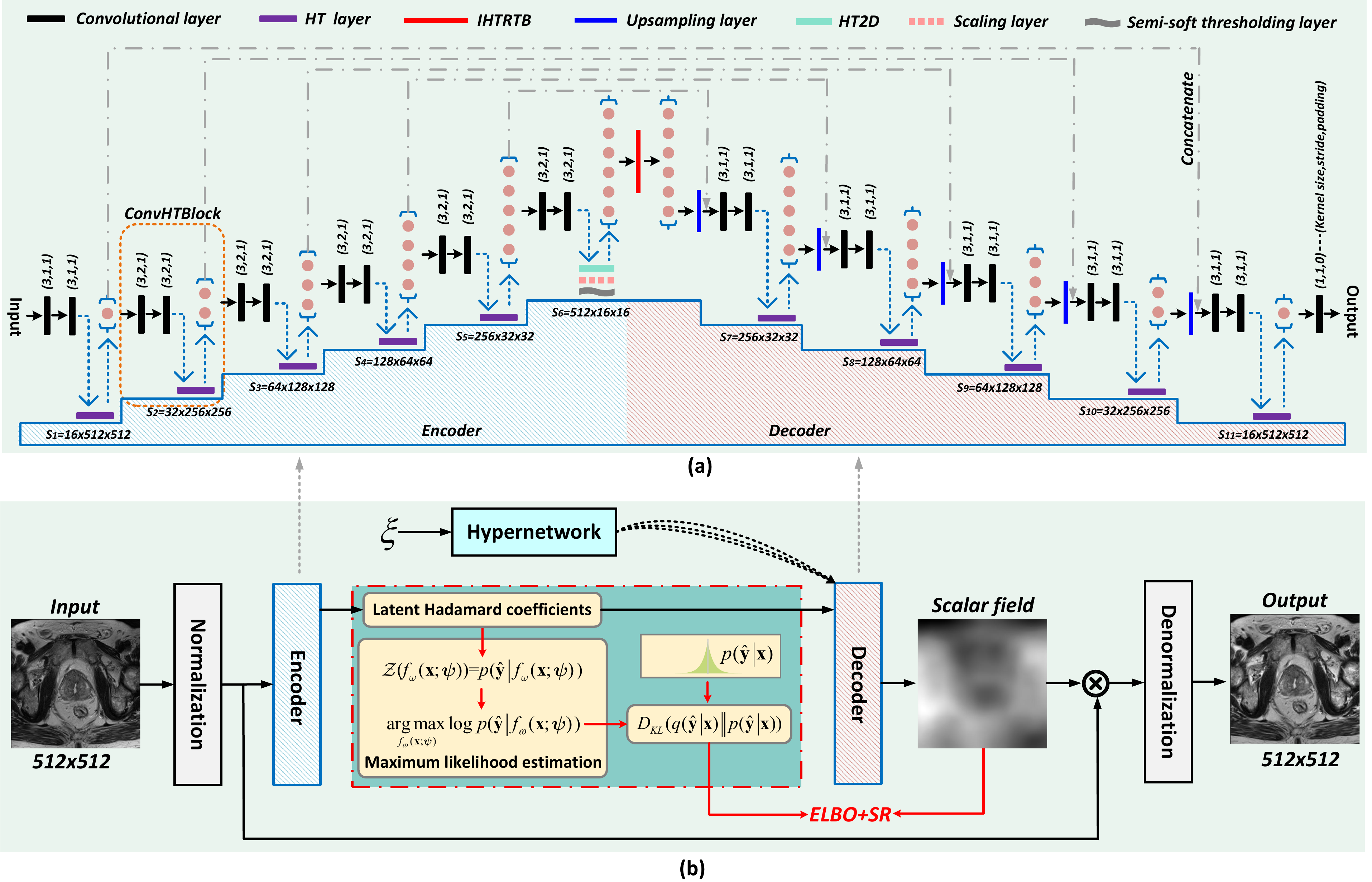}
	\caption{(a) VHU-Net architecture; (b) Overview of bias field correction workflow.}
	\label{Fig:VHUNET}
\end{figure*}

\subsection{Variational Hadamard U-Net (VHU-Net)}
To maximize the ELBO, this section presents a novel variational Hadamard U-Net (VHU-Net) that combines the U-Net structure~\cite{he2023neural} with HT2D to extract the low-frequency scalar field. It features an encoder-decoder design with symmetric skip connections, allowing for the integration of low-level and high-level features.

\subsubsection{Encoder Module of the VHU-Net}
\begin{figure}[t]
	\centering
		\includegraphics[scale=.61]{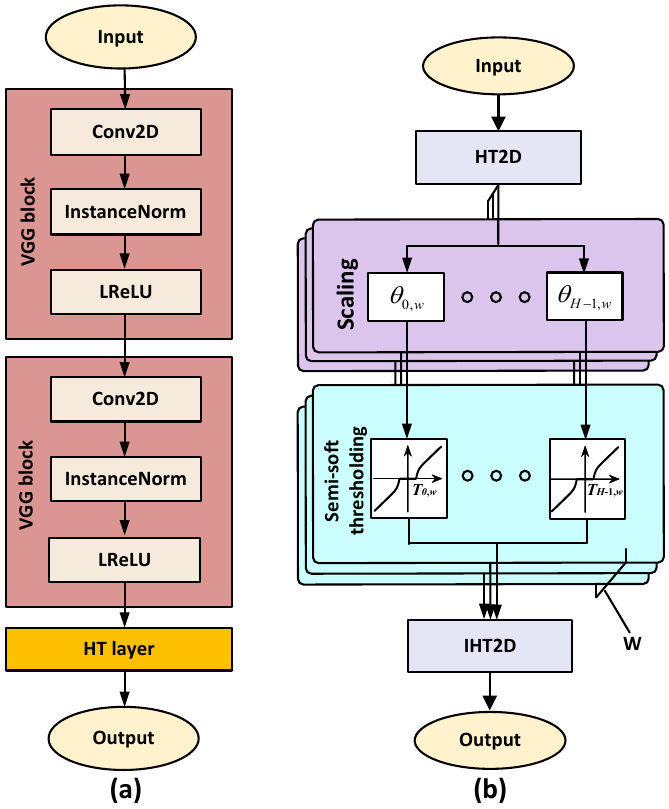}
	\caption{(a) ConvHTBlock; (b) HT layer}
	\label{Fig:ConvHTBlock}
\end{figure}

Initially, input normalization is employed as a preprocessing step to promote efficient learning~\cite{kim2025investigating}. Subsequently, the encoder of VHU-Net is designed to extract important latent space features through a hybrid framework that integrates spatial-domain convolutions with Hadamard-domain sparsification. 
As illustrated in Fig.~\ref{Fig:VHUNET} (a), the encoder comprises six hierarchical ConvHTBlocks. Each ConvHTBlock begins with two Visual Geometry Group (VGG) blocks~\cite{sinha2021completely} to extract localized features. 
As depicted in Fig.~\ref{Fig:ConvHTBlock} (a), each VGG block comprises a convolutional layer, an instance normalization layer and a leaky ReLU (LReLU) activation function, facilitating stable gradient flow and enhancing nonlinear representational capacity.

To capture bias-related frequency features, we introduce the HT layer as a frequency decomposition module within each ConvHTBlock. As shown in Fig.~\ref{Fig:ConvHTBlock} (b), the HT2D is employed to transform data $\mathbf{x}\in\mathbb{R}^{C\times H\times W}$ from the spatial domain to the Hadamard domain using Eq.~(\ref{Eq: HT}), where $C$, $H$, and $W$ denote the number of channels, the height, and the width of the tensor, respectively. 
 Subsequently, we apply two critical steps to accentuate the energy within the low-frequency band, which predominantly contains the bias field information. First, the trainable scaling layer adaptively adjusts the magnitude of frequency components, amplifying the low-frequency signals while diminishing high-frequency noise. It is computed as:
\begin{equation}
    \widetilde{\mathbf{X}} = \mathbf{\Theta}\circ {\mathbf{X}},
\end{equation}
where $\mathbf{\Theta}\in\mathbb{R}^{H\times W}$ represents the scaling parameter matrix. ${\mathbf{X}\in\mathbb{R}^{C\times H\times W}}$ denotes the output of the HT2D. An alternative interpretation of the scaling layer is its relationship with the Hadamard convolution theorem~\cite{uvsakova2002walsh,pan2024multichannel}, which posits that the element-wise product in the Hadamard domain is analogous to the dyadic convolution in the time domain:
\begin{equation}
\mathbf{u} *_d {\mathbf{v}} = \mathcal{H}^{-1}\left(\mathcal{H}(\mathbf{u})\circ\mathcal{H}(\mathbf{v})\right), 
\end{equation}
where $*_d$ denotes the dyadic convolution. $\mathbf{u}$ and $\mathbf{v}$ represent two input vectors. $\mathcal{H}(\cdot)$ stands for the HT. Hence, the scaling operation in the Hadamard domain can be converted to a convolution in the time domain to enhance feature extraction capability. However, the scaling operation introduces significantly lower computational complexity compared to convolutional filtering in the time domain~\cite{pan2023hybrid}.

\begin{figure}[t]
	\centering
		\includegraphics[scale=.7]{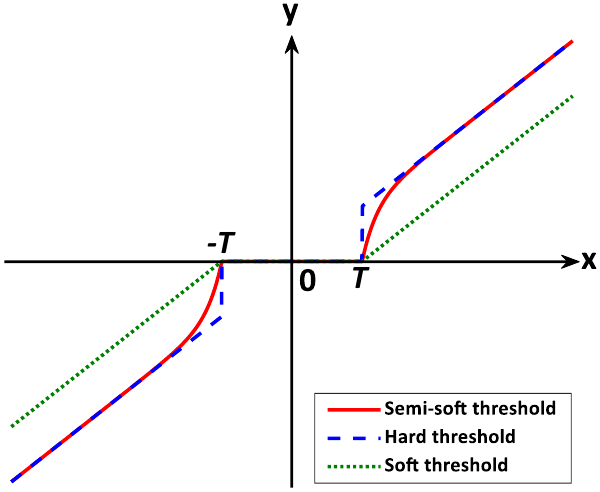}
	\caption{Visualization of threshold functions}
	\label{Fig:semi}
\end{figure}

Instead of using traditional thresholding techniques, a new trainable semi-soft thresholding layer is developed to eliminate high-frequency components. Unlike hard thresholding, which creates abrupt cutoffs that introduce ringing artifacts~\cite{panigrahi2021joint}, or soft thresholding, which uniformly shrinks large coefficients and increases estimation errors~\cite{zhu2024electroencephalogram}, the semi-soft approach provides a balanced solution. As shown in Fig.~\ref{Fig:semi}, it completely removes coefficients below the threshold while applying a graduated shrinkage to larger coefficients, maintaining the important low-frequency components that correspond to the bias field. 
The trainable semi-soft thresholding layer is derived from a binary gating operator, which is defined as follows:
\begin{equation}
\mathcal{P}({\widetilde{\mathbf{X}},\mathbf{T}}) = \text{sign}( \text{ReLU}(|\widetilde{\mathbf{X}}| - \mathbf{T}) ),
\end{equation}
where $\text{ReLU}(\cdot)$ stands for the rectified linear unit (ReLU) function~\cite{fukushima1969visual} and $\mathbf{T}\in\mathbb{R}^{H\times W}$ is non-negative trainable threshold parameters. Then, the semi-soft thresholding operator can be computed as follows:
\begin{align}
\label{Eq: SST}
\mathcal{S}(\widetilde{\mathbf{X}}, \mathbf{T}) &=\mathcal{P}({\widetilde{\mathbf{X}},\mathbf{T}})\circ\text{sign}(\widetilde{\mathbf{X}}) \circ ( |\widetilde{\mathbf{X}}| - \mathbf{T} \circ \exp({-|\widetilde{\mathbf{X}}| + \mathbf{T}}) ).
\end{align}
The detailed derivation process of Eq.~(\ref{Eq: SST}) is provided in~\ref{sec:dsst}.
After that, we perform the inverse HT2D (IHT2D) to transform the image from the Hadamard domain back to the spatial domain using Eq.~(\ref{Eq: IHT}). 
Through multiple  ConvHTBlocks, the VHU-Net processes the input and outputs the latent space representation. As shown in Fig.~\ref{Fig:VHUNET} (a), $S_i$ stands for the dimensions of the $i$-th ConvHTBlock output. At each stage, the number of spheres \( j \) corresponds to the number of feature maps \( N_e \), where $N_e = 2^{j+3}$.
Notably, the final ConvHTBlock in the encoder omits the IHT2D operation. This design choice preserves the latent representation in the Hadamard domain, allowing the decoder transformer layer to attend to frequency-decomposed features.

\begin{figure}[t]
	\centering
		\includegraphics[scale=.38]{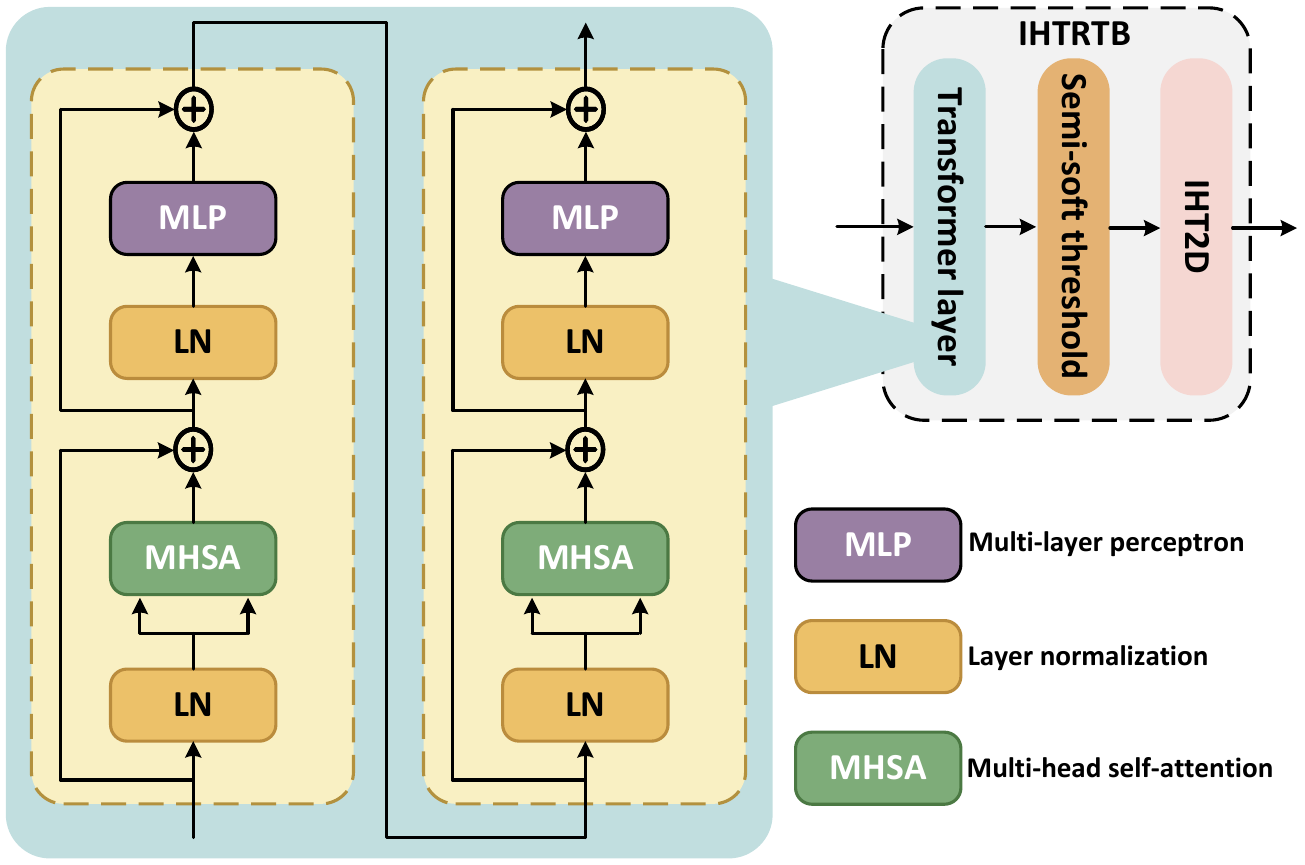}
	\caption{IHT-reconstructed transformer block.}
	\label{Fig:IHTRTB}
\end{figure}

\subsubsection{Decoder Module of the VHU-Net}
The HT layer performs channel-wise spectral transformation, enabling intra-channel frequency feature extraction. However, it inherently neglects the inter-channel dependencies that are crucial for capturing the global structure of the bias field. To overcome this limitation, an IHT-reconstructed transformer block (IHTRTB) is integrated into the decoder module to implement a global frequency-aware attention mechanism, as shown in Fig.~\ref{Fig:IHTRTB}. This design enhances the model’s sensitivity to spatially coherent low-frequency patterns associated with the bias field, while suppressing irrelevant high-frequency information across channels.

Let $\mathbf{Y}\in\mathbb{R}^{S \times M \times N}$ represent the encoded feature map, where \( S \) denotes the number of channels and \( M \times N \) defines the spatial resolution. The feature tensor is first reshaped into a sequence of spatial tokens $\mathbf{Y}_{\text{seq}} \in \mathbb{R}^{G\times S}$, where $G = M \times N$. 
This sequence is then passed through two successive transformer blocks. As shown in Fig.~\ref{Fig:IHTRTB}, each block contains a multi-head self-attention (MHSA) and a multi-layer perceptron (MLP). Both components are preceded by layer normalization and are equipped with residual connections to improve training stability and gradient propagation. In MHSA, the query, key, and value matrices are derived from linear projections:
\begin{equation}
\mathbf{Q}_l = \mathbf{Y}_{\text{seq}} \mathbf{W}^Q_l, \quad 
\mathbf{K}_l = \mathbf{Y}_{\text{seq}} \mathbf{W}^K_l, \quad 
\mathbf{V}_l = \mathbf{Y}_{\text{seq}} \mathbf{W}^V_l,
\end{equation}
where $l$ indexes the head number. \( \mathbf{W}^Q_l, \mathbf{W}^K_l, \mathbf{W}^V_l \in \mathbb{R}^{S \times D} \) are learnable projection matrices. $D=\frac{S}{L}$ is the dimensionality of each attention head. $L=8$ is the number of heads. The attention for each head is computed using the scaled dot-product formulation:
\begin{equation}
\text{head}_l=\text{Attention}(\mathbf{Q}_l, \mathbf{K}_l, \mathbf{V}_l) = 
\text{Softmax}\left( \frac{\mathbf{Q}_l \mathbf{K}_l^\top}{\sqrt{D}} \right) \mathbf{V}_l.
\end{equation}
Outputs from all \( L \) attention heads are concatenated and projected to the original channel dimension:
\begin{equation}
\widetilde{\mathbf{Y}}= 
\text{Concat}(\text{head}_1, \ldots, \text{head}_L) \mathbf{W}_o +\mathbf{b}_o+\mathbf{Y}_{\text{seq}},
\end{equation}
where $\mathbf{W}_o \in \mathbb{R}^{S \times S}$ is the output projection matrix. $\mathbf{b}_o\in\mathbb{R}^{S}$ denote trainable bias.
Next, the MLP~\cite{chen2024transunet} is employed to enhance feature extraction capacity:
\begin{equation}
\mathbf{Z} = \text{MLP}(\text{LayerNorm}(\widetilde{\mathbf{Y}}))+\widetilde{\mathbf{Y}},
\end{equation}
where $\text{LayerNorm}(\cdot)$ refers to layer normalization.
After two such transformer blocks, the sequence is reshaped back into its original spatial structure $\mathbf{Z}_{\text{out}}\in \mathbb{R}^{S \times M \times N}$.
Then, the semi-soft thresholding layer, as defined in Eq.~(\ref{Eq: SST}), is applied to remove redundant Hadamard coefficients and introduce nonlinearity after the MLP. The IHT2D is subsequently performed on the thresholded output to reconstruct the refined scalar field.

The decoder mirrors the encoder with five upsampling stages. At each decoding stage, the feature maps from the preceding layer are upsampled and concatenated with the corresponding encoder outputs via skip connections as shown in Fig.~\ref{Fig:VHUNET} (a). The merged feature undergoes a ConvHTBlock. 

During neural network training, the continuous updating of weights causes the input distributions to deeper layers to shift over time. This shift compels downstream layers to constantly readjust to the changing inputs, which can lead to issues such as gradient saturation and slower convergence. To mitigate this problem and enhance the adaptability to varying correction conditions, a hypernetwork is introduced in the decoder. This auxiliary network generates channel-wise affine parameters for each decoder ConvHTBlock, serving as a generalized form of input normalization. By stabilizing the inputs, the network becomes more robust to distributional shifts during training.
Additionally, this hypernetwork is implemented as a lightweight MLP consisting of three fully connected layers with Leaky ReLU activations. Let \( \mathbf{g}^{\ell} \in \mathbb{R}^{ C^{\ell} \times H^{\ell} \times W^{\ell}} \) represent the output feature map of the \( \ell \)-th decoder VGG block, where $C^{\ell}$, $H^{\ell}$, and $W^{\ell}$  correspond to the number of channels, height, and width, respectively. 
The modulated output \( \tilde{\mathbf{g}}^{\ell} \) is computed as:
\begin{equation}
    \tilde{\mathbf{g}}^{\ell} = \boldsymbol{\gamma}^{\ell} \circ \mathbf{g}^{\ell} + \boldsymbol{\beta}^{\ell},
\end{equation}
where \( \boldsymbol{\gamma}^{\ell}, \boldsymbol{\beta}^{\ell}\in \mathbb{R}^{C^{\ell}} \) are the channel-wise scale and bias vectors for the \( \ell \)-th decoder ConvHTBlock, respectively. All modulation vectors are generated from a shared MLP 
\( \varphi: \mathbb{R} \rightarrow \mathbb{R}^{2 \sum_{\ell=0}^{n-1} C_\ell} \), 
conditioned on a scalar control variable \( \xi \in \mathbb{R} \):
\begin{equation}
[\boldsymbol{\gamma}^0, \boldsymbol{\beta}^0, \dots, \boldsymbol{\gamma}^{n-1}, \boldsymbol{\beta}^{n-1}] = \varphi(\xi),    
\end{equation}
where $n$ is the number of decoder ConvHTBlocks.
Finally, the estimated scalar field is obtained by a \( 1 \times 1 \) convolutional layer. The bias-corrected MRI output is then computed by element-wise multiplication between the original input and the estimated scalar field, as illustrated in Fig.~\ref{Fig:VHUNET} (b). Here, the scalar field corresponds to the inverse of the bias field. A denormalization operation is ultimately applied to the bias-corrected output to restore it to the original intensity scale.

\subsection{The Evidence Lower Bound (ELBO) for VHU-Net}
\label{sec:elboc}
As established in Section~\ref{sec:formation}, our objective is to maximize the ELBO to facilitate accurate bias field estimation. This requires jointly minimizing the KL divergence term $\mathcal{D}_{KL}\left( q(\mathbf{y} | \mathbf{x}) \,\|\, p(\mathbf{y} | \mathbf{x}) \right)$ and maximizing the expected log-likelihood term $\mathbb{E}_{q(\mathbf{y} | \mathbf{x})} \left[ \log p({\mathbf{z}} | \mathbf{y}, \mathbf{x}) \right] $.
Within the proposed VHU-Net framework, $ q(\mathbf{y}|\mathbf{x})$ stands for the approximate posterior distribution of the latent Hadamard coefficients $\mathbf{y}$, given the bias-corrupted image $\mathbf{x}$. This distribution is learned via the encoder module of VHU-Net. 

Prior work on frequency-domain analysis indicates that the HT coefficients, especially the alternating current (AC) components, can be effectively modeled by Laplacian distributions~\cite{yuan1986zonal}.
Additionally, following semi-soft thresholding and filtering, highly correlated components are suppressed.
Hence, latent coefficients $y_i$ can be modeled as independent and identically distributed Laplacian variables, i.e., $y_i\sim La(\mu=0, \omega=b)$ for $1 \leq i \leq N-1$, where $\mu$ denotes the location parameter, $\omega$ represents the scale parameter and $N$ is the number of HT coefficients in the latent space. 
Assume AC coefficients $\hat{\mathbf{y}}=[y_1,\dots,y_{N-1}]$.
Accordingly, the probability density function (PDF) of the approximate posterior distribution is given by:
\begin{equation}
q(\hat{\mathbf{y}}|\mathbf{x})= \left(\frac{1}{2f_{\omega}(\mathbf{x};\psi)}\right)^{N-1} \exp\left(-\frac{||\hat{\mathbf{y}}||}{f_{\omega}(\mathbf{x};\psi)}\right),
\label{Eq: px}
\end{equation}
where $||\cdot||$ indicates the 1-norm. $f_{\omega}(\mathbf{x};\psi)$ represents the scale function parameterized by an encoder with learnable weights $\psi$. To better understand the influence of the scale parameter, we analyze the limiting behavior of $q(\hat{\mathbf{y}}|\mathbf{x})$ as $f_{\omega}(\mathbf{x};\psi)\to 0$. For the case where $||\hat{\mathbf{y}}||=0$, we have
\begin{equation}
\lim_{f_{\omega} \to 0} q(\hat{\mathbf{y}}|\mathbf{x}) = \lim_{f_{\omega}\to 0} \left( \frac{1}{2f_{\omega}(\mathbf{x};\psi)} \right)^{N-1} = \infty. 
\label{Eq: y0}
\end{equation}

For any fixed $||\hat{\mathbf{y}}||\neq 0$, by applying L’Hôpital’s Rule, we observe:
\begin{equation}
\lim_{f_\omega \to 0} q(\hat{\mathbf{y}}|\mathbf{x})
= \frac{(N-1)!}{(2||\hat{\mathbf{y}}||)^{N-1}}\lim_{f_\omega \to 0}\exp\left( -\frac{||\hat{\mathbf{y}}||}{f_\omega} \right)=0,   
\label{Eq: yn0}
\end{equation}
where $(N-1)!$ represents the factorial of $N-1$. 
According to Eq.~(\ref{Eq: y0}) and Eq.~(\ref{Eq: yn0}), as the value of $f_{\omega}(\mathbf{x};\psi)$ approaches zero, the probability density at \( y_i = 0 \) tends toward infinity, while the probability of observing $y_i \neq 0$ diminishes to zero. Therefore, setting the value of the scale function $f_{\omega}(\mathbf{x};\psi)$ close to zero can be interpreted as enforcing a sparsity constraint on the Hadamard coefficients in the latent space.

In the context of bias field correction, the bias field predominantly resides in the low-frequency components, whereas high-frequency components are largely redundant. Therefore, for an ideal deep learning model capable of recovering the ground-truth bias field, the latent representation in the Hadamard domain is expected to be sparse as shown in Fig.~\ref{Fig:timefrequency} (f). To encode this spectral sparsity in the probabilistic formulation, we assume that the true posterior distribution $p(\hat{\mathbf{y}} | \mathbf{x})$ follows a Laplace distribution with a small scale parameter $\delta$. Its PDF is defined as:
\begin{equation}
p(\hat{\mathbf{y}} | \mathbf{x})=(\frac{1}{2\delta})^{N-1} \exp\left(-\frac{||\hat{\mathbf{y}} ||}{\delta}\right).
\label{Eq: ppx}
\end{equation}

The KL divergence between the approximate posterior and true posterior is then given by:
\begin{align}
&D_{KL}(q(\hat{\mathbf{y}}|\mathbf{x}) \| p(\hat{\mathbf{y}}|\mathbf{x}))\nonumber\\ &= \int q(\hat{\mathbf{y}}|\mathbf{x})  \log \frac{q(\hat{\mathbf{y}}|\mathbf{x}) }{p(\hat{\mathbf{y}}|\mathbf{x}) } \, d\hat{\mathbf{y}} \nonumber\\ 
&=(N-1)\left(\frac{ f_{\omega}({\mathbf{x}};\psi)}{\delta} + \log \frac{\delta}{f_{\omega}({\mathbf{x}};\psi)} -1\right).
\label{Eq: DKL}
\end{align}

Nevertheless, the specific value of $f_{\omega}({\mathbf{x}};\psi)$ remains undetermined.
Subsequently, we employ maximum likelihood estimation to infer the value of scale function $f_{\omega}({\mathbf{x}};\psi)$ from the observed latent AC coefficients $\hat{\mathbf{y}} = [{y}_1, \ldots, {y}_{N-1}]$. The likelihood of the latent vector $\hat{\mathbf{y}}$ given the estimated scale $f_{\omega}({\mathbf{x}};\psi)$ can be formulated as follows:
\begin{align}\label{Eq:Lik}
   		{\rm {\mathcal{Z}}}(f_{\omega}({\mathbf{x}};\psi))=p({\hat{\mathbf{y}}}|f_{\omega}({\mathbf{x}};\psi))= \prod_{i=1}^{N-1}\frac{1}{2f_{\omega}({\mathbf{x}};\psi)} \exp\left(-\frac{|{{y}_i}|}{f_{\omega}({\mathbf{x}};\psi)}\right).
  \end{align}

Maximizing the corresponding log-likelihood yields the optimal closed-form estimate for the scale parameter:
\begin{align}
\mathop{\arg\max}\limits_{f_{\omega}({\mathbf{x}};\psi)} \log p({\hat{\mathbf{y}}}|f_{\omega}({\mathbf{x}};\psi))
=\frac{1}{N-1}\sum\limits_{i=1}^{N-1}|y_{i}|.
\label{eq:argmax}
\end{align}

Substituting Eq.~(\ref{eq:argmax}) back into Eq.~(\ref{Eq: DKL}), the KL divergence becomes:
{\small
\begin{align}
&D_{KL}
=(N-1)\left(\frac{\sum\limits_{i=1}^{N-1}|y_{i}|}{(N-1)\delta} + \log(N-1)\delta-\log{\sum\limits_{i=1}^{N-1}|y_{i}|}-1\right).
\label{Eq: NDKL}
\end{align}}
It is noticed that the minimum of $D_{{KL}}$ is attained when $\sum_{i} |{y}_{i}| = \delta(N-1)$, implying that the latent AC codes are constrained not to zero but to a controlled sparse norm.
This constraint regularizes the total latent magnitude to preserve informative frequency components relevant to the bias field, while suppressing high-frequency noise. Additionally, the direct current (DC) coefficient $y_0$ encodes the dominant low-frequency content of the bias field. Enforcing sparsity on $y_0$ may result in important information loss. Hence, no regularization is applied to the DC component.
Notably, when $\delta$ becomes very small, numerical instability may arise. To address this, the KL divergence can be symmetrically reversed to $D_{\text{KL}}(p(\hat{\mathbf{y}}|\mathbf{x})\|q(\hat{\mathbf{y}}|\mathbf{x}))$, which still encourages sparsity while mitigating overflow risks.

As outlined in Section~\ref{sec:formation}, the first term in Eq.~(\ref{eq:elbo}) represents the likelihood of reconstructing the target scalar field. Maximizing $\mathbb{E}_{q(\mathbf{y} | \mathbf{x})} \left[ \log p(\mathbf{z} | \mathbf{y}, \mathbf{x}) \right] $ can be realized by reducing the mean square error (MSE) between the estimated bias field and the ground-truth map.
However, the ground-truth bias field is typically unobservable, rendering direct supervision on the estimated bias field infeasible. To address this limitation, we optimize the network to reduce the MSE between the bias-corrected and the inhomogeneity-free images.
Therefore, the ELBO objective tailored for VHU-Net is formulated as:
\begin{align}
\mathcal{E}_{v} =-\frac{1}{H}\frac{1}{W}\sum_{h=0}^{H-1}\sum_{w=0}^{W-1} \left({e}_{h,w}-{\hat{e}}_{h,w}\right)^2 -\varepsilon D_{KL},
\label{eq: FELBO}
\end{align}
where ${e}_{h,w}$ and $\hat{e}_{h,w}$ are the bias-corrected and inhomogeneity-free outputs, respectively.
$\varepsilon$ is a weighting factor. To further enhance the spatial coherence of the estimated bias field, the final loss function incorporates a smoothness regularization (SR) term $\mathcal{R}(\mathbf{z})$ as defined in Eq.~(\ref{eq:LR}):
\begin{align}
\mathcal{L}_{\text{total}} = -\mathcal{E}_{v} + \lambda \mathcal{R}(\mathbf{z}).
\label{eq:totalloss}
\end{align}
Here, we employ the Laplacian operator~\cite{chen2021abcnet} to enforce an explicit smoothness constraint, which penalizes abrupt variations in the predicted bias field.

\section{Experiments and Results}
\subsection{Synthesis of Bias Field}
\label{sec: Synthesis of bias field}
\begin{figure}
	\centering
		\includegraphics[scale=.2]{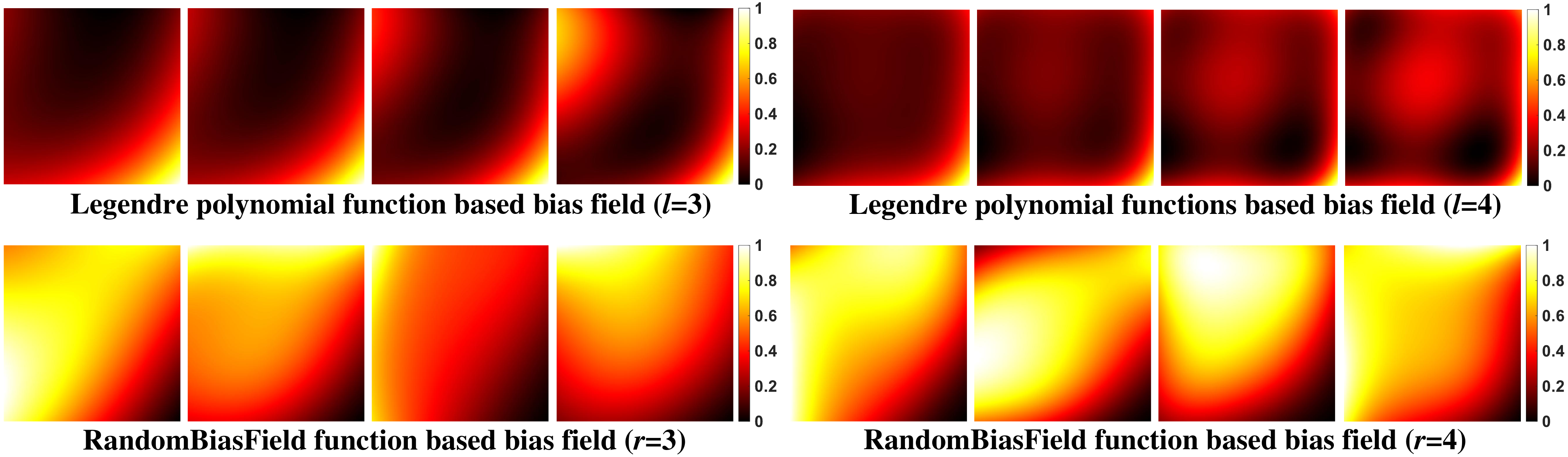}
	\caption{Simulation of MRI bias fields. The parameters $r$ and $l$ determine the spatial smoothness of the bias field.}
	\label{Fig:Simulationbias}
\end{figure}

To simulate realistic intensity inhomogeneities and assess the robustness of correction methods, we evaluated two synthetic bias field generation strategies: (1) Legendre polynomial-based functions~\cite{chen2021abcnet} and (2) the RandomBiasField function from the TorchIO library~\cite{sudre2017longitudinal}. As shown in Fig.~\ref{Fig:Simulationbias}, Legendre-based fields, constructed using polynomials up to degree $l$, produce smooth, globally varying patterns. However, these tend to lack structural diversity and are often dominated by a single low-frequency mode.
In contrast, RandomBiasField introduces spatially heterogeneous bias and yields localized low-frequency variations that more closely resemble the complex bias fields observed in clinical MRI. 
The RandomBiasField function models the MRI intensity inhomogeneity as a smooth multiplicative field generated from a linear combination of polynomial basis functions. Mathematically, for a voxel at position $(x, y, z)$, the simulated bias field $B(x, y, z)$ is defined as:
\begin{equation}
B(x, y, z) = \exp \left( 
    \sum_{i=0}^{r} 
    \sum_{j=0}^{r-i} 
    \sum_{k=0}^{r-i-j} 
    c_{ijk} \, x^{i} y^{j} z^{k}
\right),
\end{equation}
where $c_{ijk}$ are random coefficients sampled uniformly from a user-defined range $[a, b]$, controlling the amplitude of the polynomial terms. 
In bias field correction experiments, we set the parameters to $a=-0.5$ and $b=0.5$.
The parameter $r$ (referred to as \emph{order} in the implementation) determines the maximum polynomial degree and the spatial smoothness of the bias field. Smaller $r$ values yield smoother, low-frequency fields, whereas larger values introduce slightly higher spatial variation.

We further investigated field complexity by varying the order parameter $r$. 
As illustrated in Fig.~\ref{Fig:Simulationbias}, increasing $r$ from 3 to 4 resulted in richer spatial variation and greater irregularity. 
Consequently, the fields generated with $r=4$ more closely approximate real-world inhomogeneities and present greater challenges for correction compared to those with $r=3$.
Accordingly, we adopted RandomBiasField with $r=4$ in all subsequent experiments to ensure realistic and diverse simulation conditions.

\subsection{Datasets}
To evaluate the bias correction performance of the VHU-Net, we perform experiments on three representative categories of body MRI data: abdominal, prostate, and breast MRIs. For different contrasts of MRI, T2-weighted MRI offers superior contrast by emphasizing water-rich tissues, which enhances the clarity of tissue boundaries and facilitates the detection and correction of the bias field. In comparison, T1-weighted MRI exhibits lower tissue contrast and greater signal overlap, particularly with fat, making it more challenging to distinguish bias field from genuine tissue signal variations. We apply our algorithm to these different contrasts.

\subsubsection{Abdominal MRI Data}
\label{sec:abdominal}
We collected abdominal MRI images from the PanSegData dataset~\cite{zhang2025large}, comprising 15,276 real T2-weighted MRI images (382 volumes), 18,124 real T1-weighted MRI images (385 volumes), and 2,115 inhomogeneity-free T2-weighted MRI images (50 volumes), respectively. As obtaining high-quality MRI scans is extremely challenging in real-world scenarios, we adopted a limited-sample training strategy. 
Inhomogeneity-free MRI images correspond to high-quality MRI volumes manually selected by experienced radiologists from the PanSegData dataset, who are affiliated with the dataset’s original releasing laboratory. These images appear visually uniform and (nearly) free from intensity inhomogeneity. The data source and access link for this subset have been provided in the Supplementary Material for transparency and reproducibility.
For both the T1-weighted and T2-weighted modalities, the datasets were partitioned identically, with 10 volumes allocated for training and the remaining volumes reserved for testing. Specifically, the T2-weighted training set (10 volumes) comprised 305 images, whereas the testing set (372 volumes) consisted of 14,971 images. Similarly, the T1-weighted training set (10 volumes) included 328 images, and the corresponding testing set (375 volumes) contained 17,796 images.
All data splits were performed on a volume basis to avoid information leakage across slices.
To address the inherent bias field present in real MRI data, we employed the widely adopted N4ITK algorithm. The N4ITK-corrected versions of the training samples served as surrogate ground truth for supervision during training.

In addition to evaluation on the real dataset, we created a synthesized dataset for supplementary analysis. The dataset was created by multiplying inhomogeneity-free MRI images with bias fields generated using the RandomBiasField function. 
Moreover, each 3D volume is augmented by multiplying it by two distinct simulated bias fields, effectively doubling the dataset size.
The intensity range of the simulated bias fields is [0.1, 1.9], corresponding to a 90\% bias level~\cite{chen2021abcnet}.
The training set comprised 343 images (10 volumes), while the testing set included 3,887 images (90 volumes).

\begin{table}[t]
	\centering
    \small
	\caption{Imaging Protocols for Prostate Datasets.}
	\label{Tab: Acquisition}
		\begin{tabular}{lcccc} 
			\toprule
\makecell{Dataset\\(Institution)}&\makecell{Field \\strength (T)} &\makecell{Resolution (in/\\through plane) (mm)} &\makecell{Endorectal \\Coil}&{Manufacturer}\\         
			\midrule
            \midrule
			    UCL& 1.5 and 3&0.325–0.625/3–3.6 &No &Siemens \\
			    BIDMC&3 &0.25/2.2–3 &Endorectal& GE\\
                HK&1.5 &0.625/3.6 &Endorectal& Siemens\\
                RUNMC&3 &0.6–0.625/3.6–4 &Surface &Siemens \\
			\bottomrule
		\end{tabular}
\end{table}

\subsubsection{Prostate MRI Data}
For prostate MRI experiments, we employed four publicly available real T2-weighted datasets: the HK dataset~\cite{litjens2014evaluation}, the UCL dataset~\cite{litjens2014evaluation}, the RUNMC dataset~\cite{bloch2015nci}, and the BIDMC dataset~\cite{litjens2014evaluation}. 
The acquisition parameters for these four datasets are detailed in Table~\ref{Tab: Acquisition}. Similar to the abdominal MRI data preprocessing, the N4ITK was applied to establish the ``ground truth". For model training and evaluation, we used a cross-dataset approach to validate generalization performance. All models were trained on the UCL dataset, which provided a training set of 318 MRI images (13 volumes). Testing was conducted on three other datasets, comprising 288 MRI images (12 volumes) from the HK dataset, 578 MRI images (30 volumes) from the RUNMC dataset, and 521 MRI images (12 volumes) from the BIDMC dataset. 

\subsubsection{Breast MRI Data}
For the breast MRI bias field correction task, we utilized the QIN-BREAST dataset~\cite{Yankeelov2019QINBREAST02} from the cancer imaging archive (CIA), which includes 34 T1-weighted breast MRI scans. The imaging data were acquired at two institutions, Vanderbilt University Medical Center and the University of Chicago, using Philips 3T MRI scanners.
To remove residual artifacts, N4ITK correction was applied to obtain clean MRI volumes for evaluation. Additionally, a synthesized dataset was generated by multiplying clean MRI images with bias fields produced using the RandomBiasField function, following the same configuration described in Section~\ref{sec:abdominal}. The synthesized dataset was divided into a training set of 500 images (5 volumes) and a testing set of 2,900 images (29 volumes).

\subsection{Baseline Method}
To evaluate the bias correction performance of the proposed model, we compare it with N4ITK~\cite{tustison2010n4itk}, a well-established standard for bias field correction. Additionally, we extend the analysis by benchmarking the proposed model against other advanced neural network-based deep learning methods, including ABCnet~\cite{chen2021abcnet}, NPP-Net~\cite{he2023neural} and CAE~\cite{sridhara2021bias}. Furthermore, we include a comparison with Hadamard-based learning methods, such as PHU-Net~\cite{zhu2024probabilistic}, which are effective in extracting low-frequency components like the bias field. 
For reliable evaluation in the subsequent bias field correction experiments, each method is trained and tested across multiple random seeds, and the average performance along with the standard deviation is reported.
For the N4ITK baseline, we use the N4 image filter from SimpleITK to perform bias-field correction on each 3D NIfTI volume. All algorithmic parameters (including the number of iterations, convergence threshold, B-spline grid spacing, and histogram sharpening) are kept at the default SimpleITK settings. A foreground mask is obtained by applying a random threshold sampled from a uniform distribution U(0,1) to the input image, effectively excluding background regions from the bias correction process. The algorithm is executed in double precision sitkFloat64 for numerical stability.

The proposed VHU-Net was trained and evaluated on a system equipped with an Intel Ultra 9 185H processor running at 2.50 GHz and an NVIDIA GeForce RTX 4090 GPU. During training, the model was optimized using AdamW~\cite{loshchilov2017decoupled}, guided by the loss function defined in Eq.~(\ref{eq:totalloss}). All experiments were conducted using a batch size of 5 and a learning rate of 0.001.
The hyperparameter was configured with $\xi$ set to 0.1.
Besides, the KL divergence weighting was set to $\varepsilon = 0.1$, the smoothness regularization term weighting was set to $\lambda = 0.01$, and the scale parameter of the true posterior was set to $\delta = 0.00001$.
Specifically, we initialized the loss weighting coefficients $\varepsilon$ and $\lambda$ to the same order of magnitude (e.g., all set to 1) and monitored the relative magnitudes of different loss terms during training. $\varepsilon$ and $\lambda$ were then adjusted such that the gradient contributions from each loss term were of comparable scale, ensuring balanced optimization among the KL divergence, reconstruction, and smoothness regularization terms. 
Additionally, as discussed in Section~\ref{sec:elboc}, the parameter $\delta$ controls the scale of the posterior distribution and is expected to be close to zero to encode spectral sparsity in the probabilistic formulation. Therefore, we choose a small value of $0.00001$.

\subsection{Evaluations Metrics}
\begin{figure}
	\centering
		\includegraphics[scale=.19]{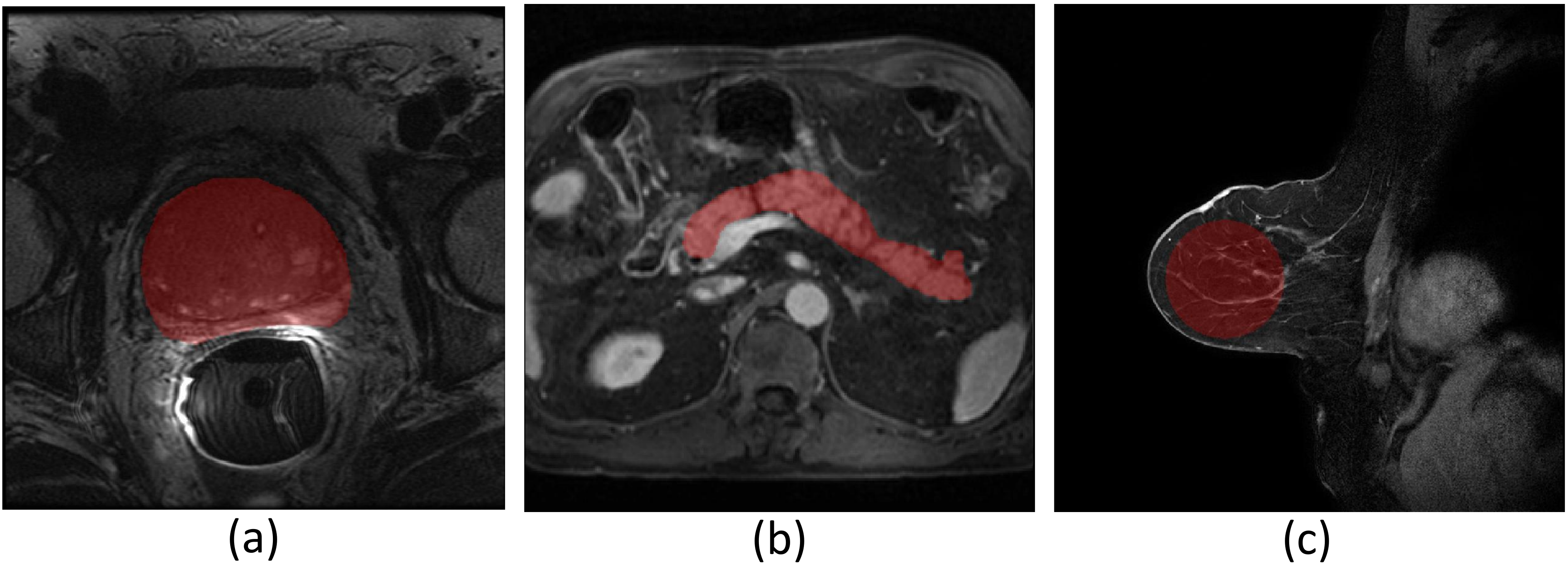}
	\caption{ROI visualizations across different MRI datasets: (a) prostate MRI, (b) abdominal MRI, and (c) breast MRI. Red regions denote ROIs.}
	\label{Fig:ROI}
\end{figure}

To comprehensively evaluate the effectiveness of the proposed bias field correction, we employ a diverse suite of quantitative metrics grouped into three categories. Intensity variation metrics include the coefficient of variation (CV)~\cite{madabhushi2005interplay}. A lower CV indicates improved tissue homogeneity. 
Its computation is defined as follows:
\begin{equation}
\mathrm{CV} = \frac{\sigma_{\mathrm{ROI}}}{\mu_{\mathrm{ROI}}},
\end{equation}
where $\mu_{\mathrm{ROI}}$ and $\sigma_{\mathrm{ROI}}$ denote the mean and standard deviation of voxel intensities within the region of interest (ROI), respectively. 
In the experimental section, ROIs are placed within a single organ or tissue type, avoiding complex anatomical structures. This prevents structural heterogeneity from biasing CV estimation. 
To clarify the ROI placement protocol, representative ROI visualizations overlaid on anatomical images for different MRI datasets are shown in Fig.~\ref{Fig:ROI}.
For datasets providing tissue annotations, ROIs are defined directly from the corresponding segmentation masks. Specifically, the prostate gland is used as the ROI in the prostate MRI dataset~\cite{litjens2014evaluation,bloch2015nci} based on the provided prostate segmentation labels, while the pancreas is selected as the ROI for abdominal MRI using the pancreas masks available in the PanSegData dataset~\cite{zhang2025large}.
In the breast MRI dataset, where no segmentation masks are available, ROIs are manually defined. In particular, a circular ROI with a fixed diameter of 120 pixels is placed within breast tissue using the brush tool in ITK-SNAP, as shown in Fig.~\ref{Fig:ROI} (c).
Additionally, the ROI is selected from regions of consistent tissue composition, ensuring that residual intensity variations primarily capture bias field effects rather than anatomical heterogeneity.
For a given subject, identical ROIs are used for all bias field correction methods to ensure a fair comparison. Besides, CV is computed on a volumetric basis and then averaged across subjects. For full reproducibility, all ROI masks used in all experiments are available at the link in the supplementary materials.

Perceptual similarity metrics assess the consistency between corrected outputs and reference data: the correlation coefficient (COCO) measures the agreement between the estimated and ground-truth bias fields, whereas structural similarity index (SSIM) and peak signal-to-noise ratio (PSNR) evaluate the pixel-level fidelity between the corrected and reference images~\cite{qiu2025memgb}. For all three, higher values indicate superior performance. 
These metrics are calculated on a per-slice (2D) basis and then averaged across slices as the method works on a pseudo-3D style (slice-by-slice).

To evaluate the impact of bias field correction on downstream tasks, we perform a segmentation experiment using corrected and uncorrected MRI data as inputs. For this task, we employ Dice score, sensitivity, and positive predictive value (PPV) as the evaluation metrics, which quantify the overlap and precision of segmentation outcomes~\cite{su2023development}.
Higher scores in these metrics demonstrate the positive impact of bias correction on downstream anatomical segmentation tasks.

\begin{table}[t]
	\centering
	\caption{Results on Synthesized PanSegData.}
	\label{Tab:simupan}
    \resizebox{\textwidth}{!}{
		\begin{tabular}{l|cccc} 
			\toprule
			\multicolumn{1}{c}{Metrics}& \multicolumn{1}{c}{SSIM ↑}& \multicolumn{1}{c}{PSNR ↑} & \multicolumn{1}{c}{COCO ↑}  &\multicolumn{1}{c}{CV $\downarrow$} 
            \\
			\midrule
            \midrule
			GT	& 1.000 	±	0.000 	& $+\infty$	&	1.000 	±	0.000 &41.480 ± 0.000  \\
			Input	&0.898 ± 0.000 & 20.632 ± 0.000 & 0.890 ± 0.000 &46.661 ± 0.000  \\
			N4ITK	&0.934 ± 0.002 & 23.817 ± 0.134 & 0.958 ± 0.002 &45.564 ± 0.017 \\
			NPP-Net &0.922 ± 0.010 & 21.687 ± 0.744 & 0.923 ± 0.004 &42.380 ± 0.540  \\
			CAE	    &0.945 ± 0.001 & 25.790 ± 0.306 & 0.964 ± 0.003 &42.209 ± 0.664 \\
            ABCnet  &0.935 ± 0.001 &23.039 ± 0.120 &0.948 ± 0.005 &43.143 ± 1.007 \\
			PHU-Net &0.928 ± 0.005 & 23.250 ± 0.514 & 0.942 ± 0.008 &41.912 ± 0.496 \\
            VHU-Net &\textbf{0.964 ± 0.000} & \textbf{26.229 ± 0.212} & \textbf{0.987 ± 0.001} & \textbf{41.695 ± 0.238} \\
			\bottomrule
            
		\end{tabular}
	}
\end{table}

\subsection{Results on Synthesized PanSegData Dataset}
\subsubsection{Quantitative Comparison}
Table~\ref{Tab:simupan} summarizes the quantitative performance of different bias field correction methods on T2-weighted abdominal MRI images with synthetic bias fields. Among all compared methods, the proposed VHU-Net achieves the best overall performance across all metrics. 
Specifically, it achieves higher SSIM and PSNR than CAE and NPP-Net, indicating better structural preservation and stronger alignment with the true image intensities. 
This performance gain arises from a key architectural insight: 
ABCnet, CAE, and NPP-Net primarily depend on convolutional layers to estimate the bias field, limiting their capacity to capture complex patterns. In contrast, VHU-Net enhances this process by combining convolutional layers with a HT layer. This hybrid architecture allows it to extract features across both temporal and frequency domains, leading to more comprehensive bias field modeling.
Additionally,  the VHU-Net supports the joint optimization of both convolutional filters and threshold parameters. This strategy enhances data adaptability and bias isolation capability.
Therefore, VHU-Net outperforms ABCnet, CAE, and NPP-Net.
In terms of average COCO, VHU-Net reaches 0.987, closely approximating the ground-truth bias field and outperforming PHU-Net. This improvement is primarily attributed to the incorporation of a transformer module within the VHU-Net architecture. It enables the model to adaptively attend to informative frequency bands within the Hadamard domain, with particular emphasis on low-frequency regions where bias field components predominantly reside, thereby enhancing estimation accuracy.

Compared to N4ITK, VHU-Net improves intensity homogeneity by reducing the average CV score from 45.564 to 41.695, reflecting a 8.49\% enhancement.
This improvement stems from the application of multiple trainable ConvHTBlocks, which encourages VHU-Net to tailor its correction strategy to varying anatomical structures and imaging conditions.
These results confirm that VHU-Net offers state-of-the-art performance in bias field correction under simulated conditions with known ground truth.

\begin{figure*}[t]
	\centering
		\includegraphics[scale=.21]{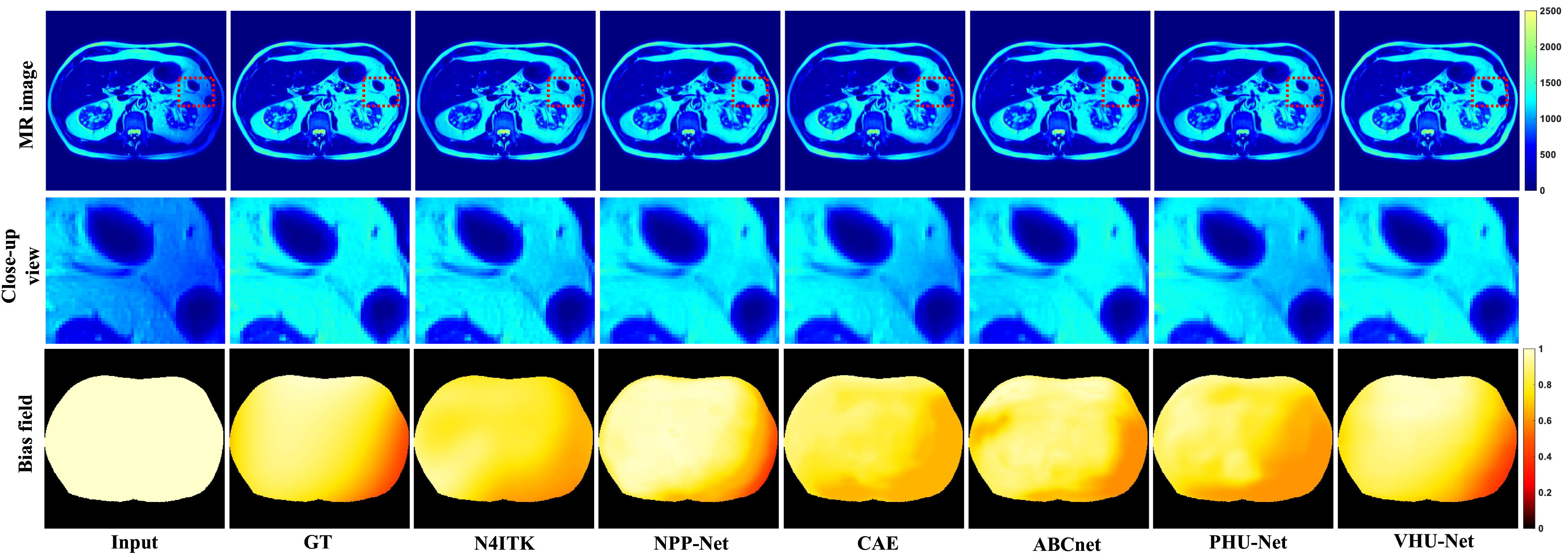}
	\caption{The qualitative comparison on synthesized PanSegData. From top to bottom, the rows show the corrected MRI images, the zoomed-in local views, and the estimated bias fields.}
	\label{Fig:simpan}
\end{figure*}

\subsubsection{Qualitative Comparison}
Fig.~\ref{Fig:simpan} provides a qualitative comparison of bias field correction methods applied to MRI data with synthetic bias fields. The figure is structured in three rows: the corrected MRI images (top), zoomed-in views for local evaluation (middle), and the estimated bias fields (bottom). 
All MRI images are displayed using pseudo-color mapping to facilitate visual comparison across different methods. Additionally, bias fields are normalized to a consistent intensity scale, enabling clearer assessment of intensity variation and smoothness.
In the top row, the corrected MRI images show that VHU-Net achieves superior bias removal, restoring tissue intensities to a visually uniform appearance while preserving anatomical detail. In contrast, CAE and PHU-Net leave noticeable residual inhomogeneities, especially in the right half of the regions. N4ITK introduces artificial intensity transitions near tissue boundaries. 
NPP-Net shows improvements over traditional methods but still fails to achieve local consistency comparable to VHU-Net.
The second row provides a zoomed-in view of a selected region, enabling finer inspection of local intensity uniformity. VHU-Net exhibits the most consistent intensity distribution within small anatomical structures. Competing methods exhibit varying degrees of local inconsistency, which can negatively impact downstream tasks such as segmentation or quantitative analysis.
The bottom row displays the estimated bias fields. Since the ground-truth bias is available in this simulation, qualitative comparison reveals that VHU-Net produces the most accurate and spatially smooth estimation, closely matching the ground-truth (GT) bias pattern. In comparison, CAE, ABCnet, PHU-Net, NPP-Net, and N4ITK fail to capture spatial variations. Enlarged and higher-resolution images are provided in the supplementary materials.

\subsection{Results on Real PanSegData Dataset}
To comprehensively validate the effectiveness of VHU-Net in real-world clinical scenarios, we applied it to real abdominal MRI images. Unlike the synthesized dataset, the real images do not have ground-truth bias fields, which limits the range of evaluable metrics. Therefore, we employed the CV as a quantitative measure of bias correction performance.

\begin{table}[htbp]
    \centering
	\caption{Results on Real PanSegData.}
	\label{Tab:realpana}
		\begin{tabular}{lcc} 
			\toprule
{Algorithms}&{T1-weighted} &{T2-weighted} \\

			\midrule
            \midrule
			Input	&27.929 ± 0.000 &46.107 ± 0.000   \\
			N4ITK	&27.310 ± 0.001 &44.846 ± 0.006   \\
			NPP-Net	&27.140 ± 0.381 &43.560 ± 0.089   \\
			CAE	    &26.727 ± 0.083 &42.597 ± 0.060   \\
            ABCnet  &26.674 ± 0.042 &42.613 ± 0.451   \\
			PHU-Net &26.821 ± 0.106 &42.967 ± 0.390   \\
            VHU-Net &\textbf{26.537 ± 0.067} &\textbf{41.615 ± 0.035}   \\
			\bottomrule
		\end{tabular}
\end{table}



\subsubsection{Quantitative Comparison}
As shown in Table~\ref{Tab:realpana}, VHU-Net again outperforms all other methods in all reported metrics on the real T2-weighted PanSegData. 
Specifically, VHU-Net reduces the CV by 7.20\%, 2.34\%, 4.47\%, 2.30\%, and 3.15\% when compared to N4ITK, ABCnet, NPP-Net, CAE, and PHU-Net, respectively. 
These results demonstrate that VHU-Net not only achieves strong performance on synthetic data but also generalizes effectively to real clinical inputs. 
Although all neural network-based methods are trained using partially N4ITK-corrected images as reference labels, VHU-Net consistently outperforms N4ITK, CAE, NPP-Net, and PHU-Net on the testing dataset. This improvement can be attributed to fundamental architectural distinctions. Specifically, CAE and NPP-Net utilize deep convolutional architectures with numerous layers to estimate the bias field, which increases the risk of overfitting. In contrast, VHU-Net incorporates the ELBO to enforce sparsity in the Hadamard domain during the training process. The sparsity constraint facilitates the activation of a minimal subset of critical neurons, preventing the model from overfitting to redundant details. This promotes more efficient learning, ultimately improving generalization ability.

To verify that our model does not hallucinate a bias field in homogeneous MRI data, we conduct an additional evaluation on ten high-quality MRI volumes that are manually selected by a senior radiologist from the PanSegData dataset. These volumes exhibit uniform intensity distribution and are regarded as bias-free for evaluation. When processed by our model, the corrected images show no visible intensity distortion or shading artifacts. Quantitatively, the average CV measures 36.312 for the input images and 35.989 after correction, yielding a relative difference of only 0.89\%. Such a small deviation demonstrates that the model remains stable when processing clean data and does not generate artificial bias fields. Visual assessment also confirms that tissue uniformity and anatomical contrast are fully preserved, as shown in Fig. 9 in the supplementary material.


Table~\ref{Tab:realpana} also summarizes the quantitative evaluation results on the T1-weighted PanSegData. 
Among all competing methods, the proposed VHU-Net attains the lowest CV (26.537 ± 0.067), suggesting that it most effectively reduces intensity non-uniformity. These results collectively demonstrate that VHU-Net provides superior performance in T1-weighted MRI inhomogeneity correction, thereby generating more consistent and diagnostically useful MR images for downstream segmentation tasks.

\begin{figure}[htbp]
	\centering
		\includegraphics[scale=.23]{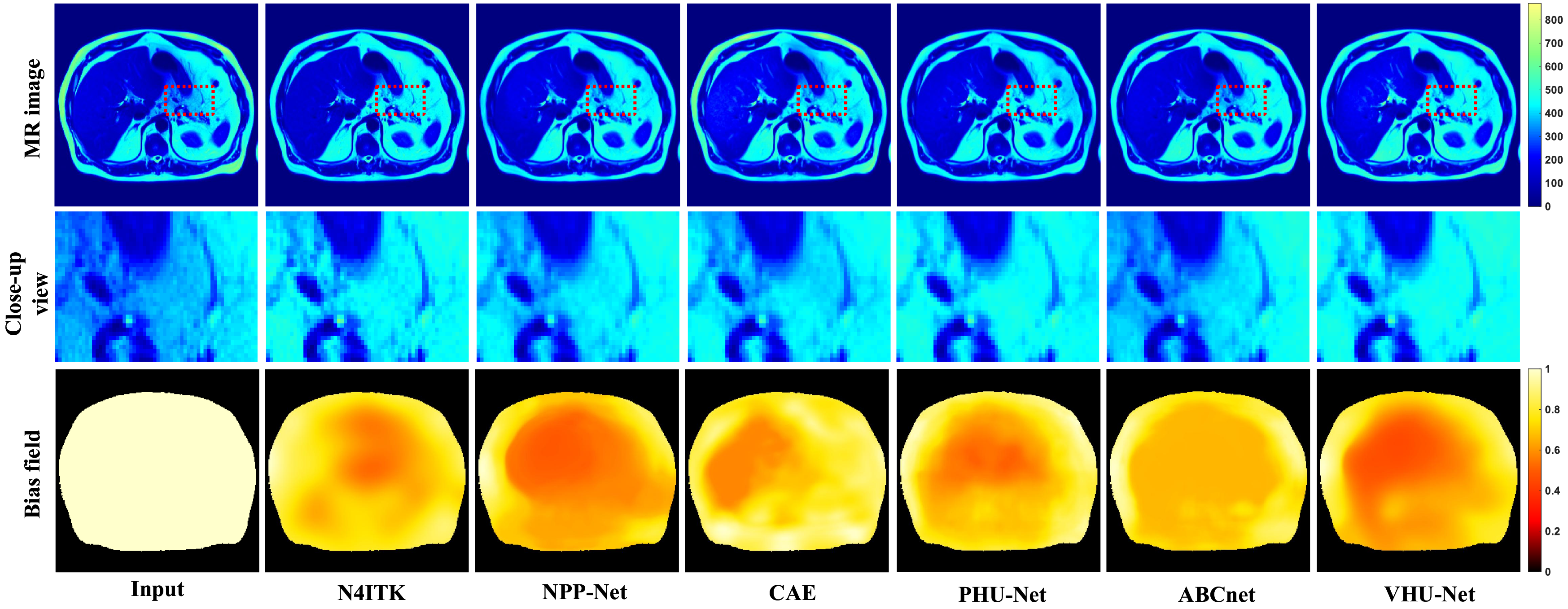}
	\caption{The qualitative comparison on real T2-weighted PanSegData. From top to bottom, the rows show the corrected MRI images, the zoomed-in local views, and the estimated bias fields.}
	\label{Fig:realpan}
\end{figure}

\subsubsection{Qualitative Comparison}
Fig.~\ref{Fig:realpan} illustrates the qualitative performance of various bias field correction methods on the T2-weighted abdominal dataset. 
The uncorrected input image exhibits visible intensity inhomogeneities, especially around the organ boundaries, which compromise tissue contrast and anatomical interpretability. 
N4ITK, NPP-Net, CAE, ABCnet, and PHU-Net produce smooth bias fields and enhance intensity uniformity. However, these methods still leave residual artifacts or compromise anatomical detail, as shown in the full and zoomed-in views.
In comparison, the proposed VHU-Net demonstrates the best overall performance. The corrected images exhibit high intensity homogeneity, enhanced tissue contrast, and well-preserved anatomical details. Notably, the estimated bias field is smooth, spatially coherent, and consistent with expected low-frequency characteristics.
These qualitative observations suggest that VHU-Net exhibits robust performance on real clinical data.

\begin{figure}[t]
	\centering
		\includegraphics[scale=.23]{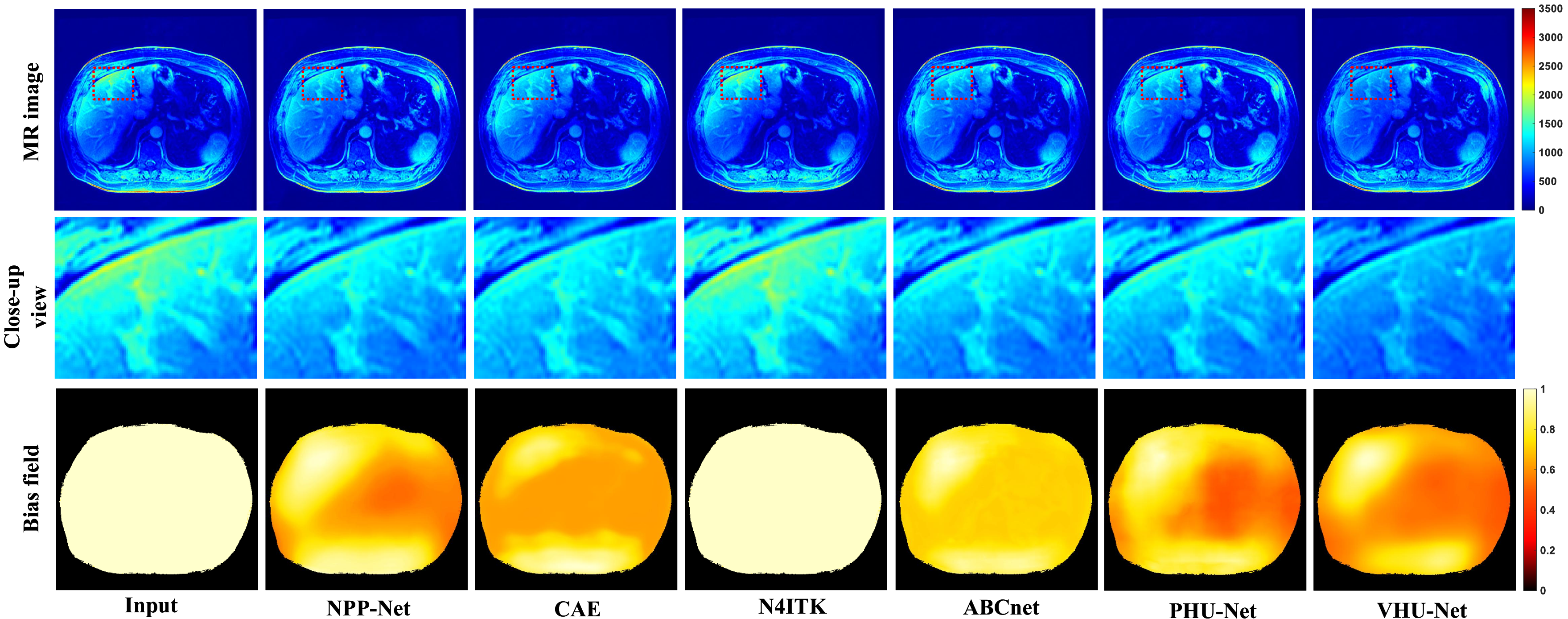}
	\caption{The qualitative comparison on real T1-weighted PanSegData. From top to bottom, the rows show the corrected MRI images, the zoomed-in local views, and the estimated bias fields.}
	\label{Fig:realpanT1}
\end{figure}

Fig.~\ref{Fig:realpanT1} presents the qualitative comparison of different bias field correction methods on the real T1-weighted PanSegData. The original MR image exhibits pronounced intensity inhomogeneity, particularly in peripheral regions. The result obtained by N4ITK appears visually similar to the uncorrected input, indicating that the correction was largely ineffective. This failure stems from the fact that N4ITK assumes intensity uniformity across distinct tissue types. However, this assumption does not hold for T1-weighted abdominal MRI, where complex anatomical structures and heterogeneous tissue compositions cause substantial natural intensity variations. Consequently, N4ITK struggles to distinguish between genuine anatomical contrast and bias-related shading, leading to under-correction. Other learning-based methods demonstrate more effective bias removal and yield improved visual homogeneity. Nevertheless, their corrected images still show local intensity inconsistencies, especially in regions with strong field distortion, as illustrated in the close-up view. In contrast, the proposed VHU-Net provides the most visually consistent result, producing spatially coherent bias field estimates while preserving anatomical detail. These observations suggest that VHU-Net better adapts to the complex signal distribution of real abdominal MR images, effectively separating bias artifacts from inherent tissue contrast.

\subsection{Results on Multi-center Prostate Dataset}
To assess the cross-domain generalizability of the proposed method, we conducted experiments on multi-center prostate MRI datasets.

\begin{table}[t]
	\centering
	\caption{Results on Multi-center Prostate MRI Dataset. }
	\label{Tab:realpros}
		\begin{tabular}{lccc} 
			\toprule
{Data}&{HK} &{BIDMC} &{RUNMC} \\             
			\midrule
            \midrule
			Input	&43.058 ± 0.000 &58.064 ± 0.000 &32.567 ± 0.000 \\
			N4ITK	&40.900 ± 1.526 &53.351 ± 3.333 &33.821 ± 0.802 \\
			NPP-Net	&34.038 ± 0.884 &43.131 ± 0.528 &30.753 ± 0.623 \\
			CAE	    &37.801 ± 0.863 & 50.821 ± 0.354 & 30.845 ± 0.108\\
            ABCnet  &34.037 ± 0.137 &44.325 ± 1.021 &31.012 ± 0.224 \\
			PHU-Net &35.139 ± 0.323 & 43.320 ± 0.323 & 30.830 ± 0.081\\
            VHU-Net &\textbf{33.864 ± 0.185} & \textbf{41.959 ± 0.752} & \textbf{30.695 ± 0.038}\\
			\bottomrule
		\end{tabular}
\end{table}

\subsubsection{Quantitative Comparison}
 Quantitative results are summarized in Table~\ref{Tab:realpros}. Compared to other methods, VHU-Net delivers competitive results in terms of intensity uniformity. Specifically, on the HK dataset, it achieves the lowest CV. Similar trends are observed on the BIDMC and RUNMC datasets, where VHU-Net outperforms both conventional methods (e.g., N4ITK) and learning-based approaches (e.g., ABCnet, NPP-Net), demonstrating superior bias field correction and improved image quality.
Therefore, VHU-Net maintains robust performance across site-specific imaging variations, highlighting its resilience to domain shifts.
 These results confirm that VHU-Net is not only effective in controlled settings but also scales well across heterogeneous clinical datasets, making it suitable for real-world deployment. Besides, as shown in Table~\ref{Tab: Acquisition}, these datasets have significant differences in: Field Strength (e.g., 1.5T vs. 3T), Coil Configuration (e.g., endorectal), and Spatial Resolution (e.g., voxel spacing and slice thickness). These technical differences in hardware and protocol are the direct cause of the variations observed in the CV values between the three datasets.

\begin{figure*}[t]
	\centering
		\includegraphics[scale=.23]{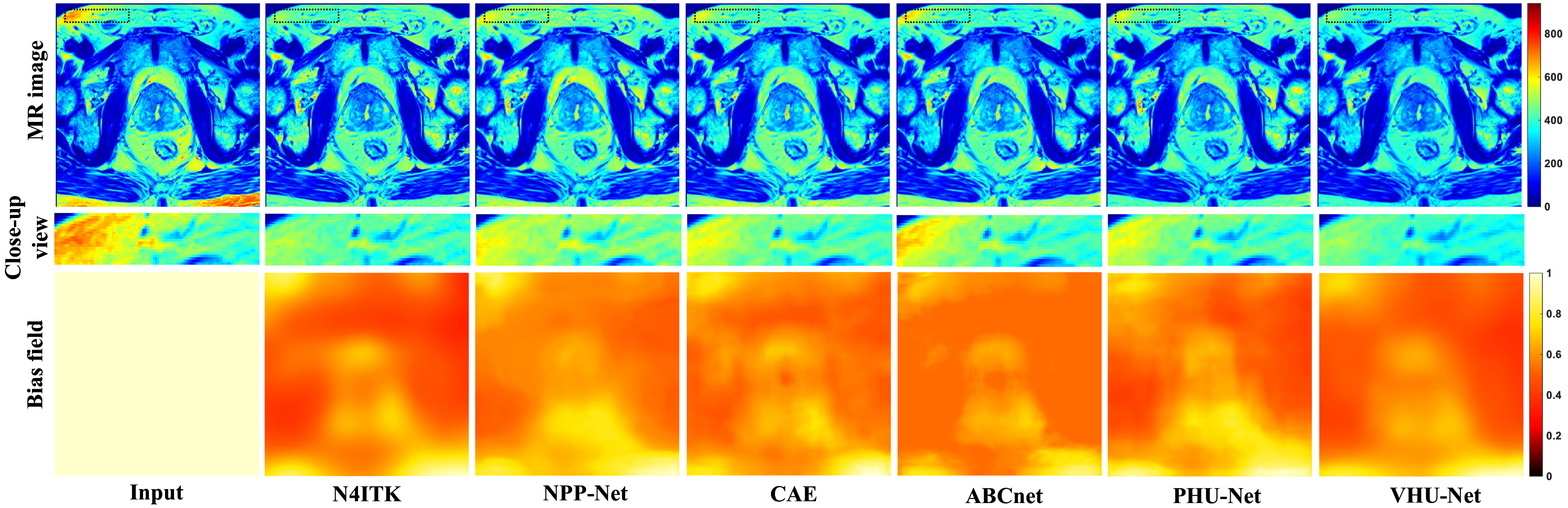}
	\caption{The qualitative comparison on prostate dataset. From top to bottom, the rows show the corrected MRI images, the zoomed-in local views, and the estimated bias fields.}
	\label{Fig:realpros}
\end{figure*}

\subsubsection{Qualitative Comparison}
Fig.~\ref{Fig:realpros} provides a comparative visualization of bias field correction performance across several representative methods on real prostate MRI images. While NPP-Net, CAE, ABCnet, and PHU-Net generally generate smooth bias fields, they exhibit insufficient compensation near the margins, leaving edge-related artifacts unaddressed. 
N4ITK demonstrates strong global bias correction capabilities. however, it tends to over-brighten the region above the prostate, resulting in the loss of fine anatomical details.
The proposed VHU-Net addresses these challenges effectively. 
The restored image exhibits improved uniformity and structural detail, without introducing over-smoothing or noise artifacts. Besides, it produces a bias field that is both smooth and anatomically consistent.

\subsection{Results on Breast MRI Dataset}
To evaluate the robustness and cross-anatomy generalizability of our proposed method, we further validate it using synthesized breast MRI datasets.

\subsubsection{Quantitative Comparison}

\begin{table}[t]
    \centering
    \caption{Results on T1-weighted Breast MRI Dataset.}
	\label{Tab:simubrea}
    \resizebox{\textwidth}{!}{
		\begin{tabular}{l|cccc} 
			\toprule
			\multicolumn{1}{c}{Metrics}& \multicolumn{1}{c}{SSIM ↑}& \multicolumn{1}{c}{PSNR ↑} & \multicolumn{1}{c}{COCO ↑}  &\multicolumn{1}{c}{CV $\downarrow$} 
            
            \\
			\midrule
            \midrule
			GT	& 1.000 	±	0.000 	& $+\infty$	&	1.000 ±	0.000 &50.030 ± 0.000 \\
			Input	&0.918 ± 0.000 & 28.637 ± 0.000 & 0.885 ± 0.000 &63.116 ± 0.000  \\
			N4ITK	&0.948 ± 0.000 & 29.951 ± 0.000 & 0.955 ± 0.000 &54.871 ± 0.001  \\
			NPP-Net &0.950 ± 0.002 & 30.473 ± 0.252 & 0.970 ± 0.001 &57.464 ± 0.339\\
			CAE	    &0.937 ± 0.001 & 29.301 ± 0.053 & 0.926 ± 0.002 &58.661 ± 0.152  \\
			PHU-Net &0.953 ± 0.004 & 30.179 ± 0.466 & 0.971 ± 0.006 &57.053 ± 1.155  \\
            ABCnet &0.937 ± 0.001 & 29.293 ± 0.056 & 0.904 ± 0.007 &58.570 ± 0.558 \\    
                VHU-Net &\textbf{0.964 ± 0.003} & \textbf{31.079 ± 0.479} & \textbf{0.985 ± 0.001} &\textbf{54.669 ± 0.317}\\
			\bottomrule
            
		\end{tabular}
	}
\end{table}

            
            

Table~\ref{Tab:simubrea} summarizes the quantitative performance of different methods on the T1-weighted breast MRI dataset. Compared with the uncorrected input, all correction algorithms improve both SSIM and PSNR, indicating effective suppression of intensity bias and recovery of image fidelity.
Among all competing methods, VHU-Net achieves the best overall performance. It records the highest SSIM and PSNR, demonstrating superior preservation of structural details and reduced reconstruction error. 
Furthermore, the COCO score of VHU-Net surpasses all other methods. It indicates that the estimated bias field most closely matches the ground truth, reflecting highly accurate bias field reconstruction.
The lowest CV value of VHU-Net (54.669 ± 0.317) confirms that the method effectively suppresses intensity bias.
Overall, these results demonstrate that VHU-Net consistently provides the best trade-off between homogeneity correction and structure preservation, closely approximating the ground truth image quality on synthesized data.

\subsubsection{Qualitative Comparison}

\begin{figure}[htbp]
	\centering
		\includegraphics[scale=.21]{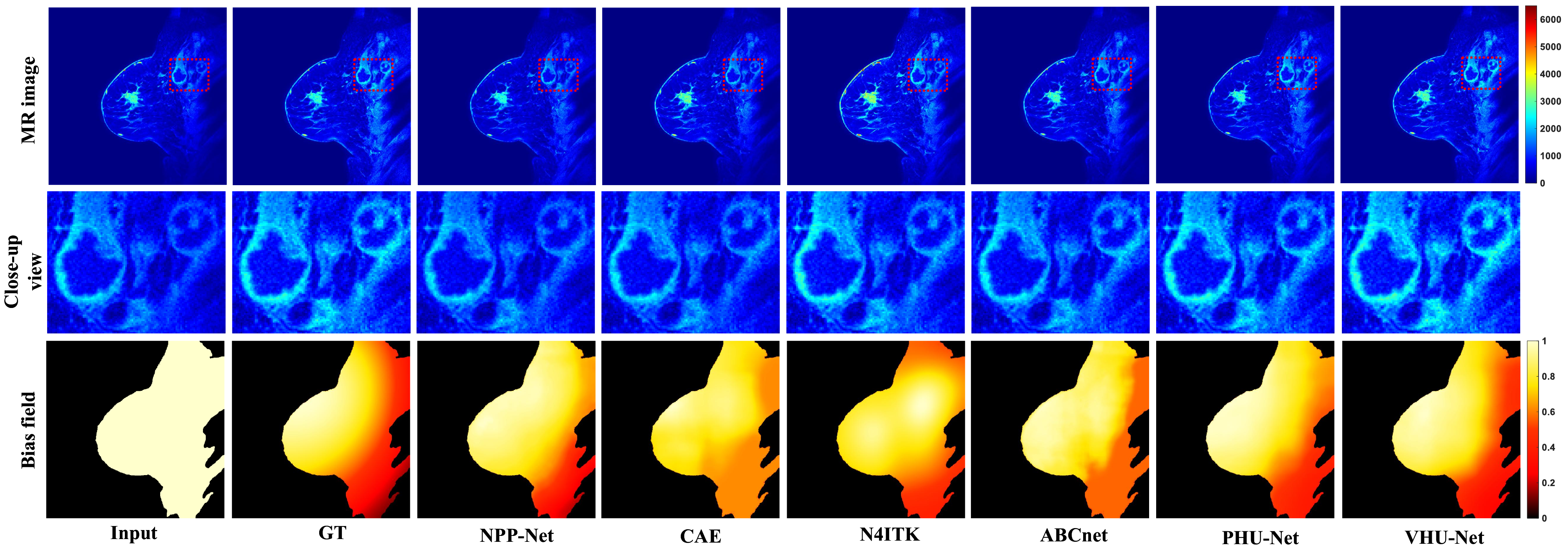}
	\caption{The qualitative comparison on T1-weighted breast dataset. From top to bottom, the rows show the corrected MRI images, the zoomed-in local views, and the estimated bias fields.}
	\label{Fig:realbreT1}
\end{figure}

As shown in Fig.~\ref{Fig:realbreT1}, the visual comparison further supports the quantitative findings. The uncorrected MR image exhibits clear intensity inhomogeneity, while N4ITK and CAE produce only partial correction and leave residual shading. Learning-based methods such as NPP-Net, ABCnet, and PHU-Net generate more uniform intensity but still fail to fully recover the true contrast distribution. In contrast, VHU-Net yields the most visually consistent result, with homogeneous intensity and well-preserved structural details. The estimated bias field from VHU-Net is also the closest to the true bias field, confirming its accurate bias estimation and reliable correction capability.

\begin{figure}[t]
	\centering
		\includegraphics[scale=0.56]{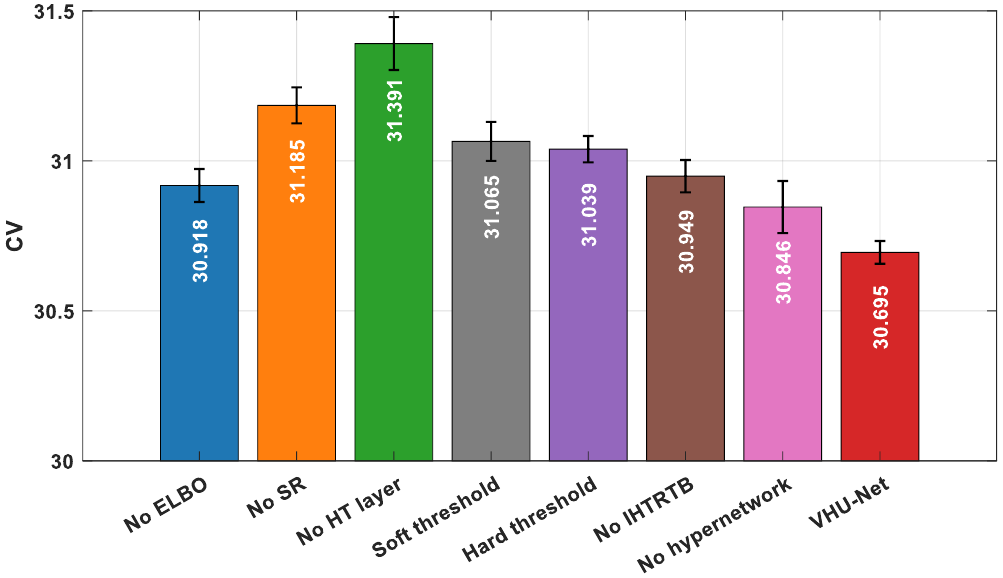}
	\caption{Ablation experiment results.}
	\label{Fig:ablation}
\end{figure}

\subsection{Ablation Study}
To validate the effectiveness of each module in the proposed VHU-Net, we performed a comprehensive ablation study on the RUNMC dataset, as summarized in Fig.~\ref{Fig:ablation}.  

\subsubsection{Ablation Study on HT Layer}
When the HT layer is not present in the VHU-Net, the performance degrades, as a higher CV is obtained.  It is because the HT decomposes image signals into orthogonal frequency components, allowing the network to separate and model the low-frequency bias field more effectively. Additionally, the scaling operation in the Hadamard domain is similar to the convolution operation in the spatial domain. This enhances the capability to extract key features for bias field estimation.

\subsubsection{Ablation Study on IHTRTB}
Removing the IHTRTB module from VHU-Net leads to a noticeable degradation in correction performance. Specifically, the average CV increases from 30.695 to 30.949, reflecting a 0.83\% rise. 
This is because the IHTRTB operates as a global frequency-aware attention mechanism that effectively captures inter-channel dependencies. It emphasizes informative frequency bands where bias fields are concentrated, thus promoting accurate and spatially coherent bias field reconstruction.

\subsubsection{Ablation Study on Smoothness Regularization}
 When VHU-Net is trained without smoothness regularization (SR), the model experiences a measurable decline in performance. 
This is due to the fact that the regularization term plays an essential role in promoting spatial smoothness and penalizing abrupt variations in the predicted bias field, thereby improving bias field estimation accuracy.

\subsubsection{Ablation Study on ELBO}
To assess the contribution of the ELBO in the proposed framework, we performed an ablation study by substituting the ELBO objective with the conventional MSE loss. As shown in Fig.~\ref{Fig:ablation}, in the absence of ELBO, the CV increased from 30.695 to 30.918. 
By enforcing sparsity and effective regularization in the latent space, the ELBO encourages the elimination of redundant coefficients, such as high-frequency noise and irrelevant details. This facilitates stable optimization and enables more accurate modeling of the underlying bias field.

\subsubsection{Ablation Study on Thresholding Layer}
Compared to VHU-Net variants incorporating soft-thresholding and hard-thresholding layers, the VHU-Net equipped with a semi-soft-thresholding layer achieves the lowest CV.
Although the soft-thresholding function removes small high-frequency components, it also attenuates large low-frequency coefficients, leading to errors in bias field estimation. Due to the discontinuity at the threshold, hard-thresholding causes abrupt truncation of frequency components. These sharp spectral transitions introduce discontinuities that manifest in the spatial domain as ringing artifacts~\cite{panigrahi2021joint}, ultimately degrading bias field correction quality. However, the semi-soft threshold function eliminates the abrupt discontinuities introduced by the hard-threshold function and alleviates the constant bias issue associated with the soft-threshold function, as shown in Fig.~\ref{Fig:semi}.

\subsubsection{Ablation Study on Hypernetwork}
To evaluate the impact of the hypernetwork, we exclude it from the VHU-Net. This modification results in a deterioration in the quantitative performance metric, as indicated by an increase in CV.
The reason is that the hypernetwork is responsible for generating adaptive scale and bias parameters for each decoder block. It enhances the capacity of the VHU-Net to accommodate distributional shifts during training and to adapt to different correction conditions.

\subsection{Segmentation Experiment}

Bias field correction aims to support and optimize subsequent medical imaging workflows. In this section, prostate and pancreas segmentation experiments are used to verify the validity of different correction models. The process begins by applying a pre-trained bias field correction module to the input medical images. The corrected images are then passed to the segmentation model, which has been trained specifically for each target organ. 
In this study, we utilize the well-established TransUNet~\cite{chen2024transunet} and nnUNetV2~\cite{isensee2024nnu} for segmentation, given their demonstrated superior performance across diverse medical image segmentation tasks.
For prostate segmentation, the model is pre-trained on a public multi-center dataset that includes BIDMC, I2CVB and UCL datasets~\cite{liu2020shape}, and evaluated on the HK dataset~\cite{litjens2014evaluation} after applying various bias correction methods. Similarly, for pancreas segmentation, the model is trained on the real PanSegData dataset and evaluated on the simulated PanSegData dataset as mentioned in Section~\ref{sec:abdominal}. In the above segmentation experiments, all the training and testing datasets were constructed with no overlapping samples.
To ensure consistency and fairness in comparison, all experiments adhere to the same training configurations and evaluation protocols established in the prior benchmark study~\cite{chen2021transunet}.




\begin{table}[htbp]
	\centering
	\caption{TransUnet Prostate Segmentation Results.}
	\label{Tab:transpro}
		\begin{tabular}{lccc} 
			\toprule
{Metrics}&{Dice $\uparrow$} &{Sensitivity $\uparrow$} &{PPV $\uparrow$} \\             
			\midrule
            \midrule
			Input	&0.686 ± 0.138  &0.592 ± 0.159 &0.856 ± 0.085   \\
            Augmented  &0.692 ± 0.133 & 0.603 ± 0.155 & 0.850 ± 0.088\\
			N4ITK	&0.697 ± 0.125  &0.606 ± 0.148  &0.852 ± 0.088  \\
			NPP-Net	&0.742 ± 0.070  &0.660 ± 0.096  &\textbf{0.859 ± 0.078}  \\
			CAE	    &0.702 ± 0.164  &0.620 ± 0.174  &0.844 ± 0.101  \\
            ABCnet  &0.721 ± 0.100  &0.637 ± 0.127  &0.850 ± 0.085 \\
			PHU-Net &0.694 ± 0.127  &0.600 ± 0.150  &0.858 ± 0.084  \\
                VHU-Net &\textbf{0.756 ± 0.063}  &\textbf{0.684 ± 0.090}  &0.857 ± 0.074  \\
			\bottomrule
		\end{tabular}
\end{table}

\begin{table}[htbp]
	\centering
	\caption{TransUnet Pancreas Segmentation Results.}
	\label{Tab:transpan}
		\begin{tabular}{lccc} 
			\toprule
{Metrics}&{Dice $\uparrow$} &{Sensitivity $\uparrow$} &{PPV $\uparrow$} \\             
			\midrule
            \midrule
			Input	&0.710 ± 0.063 & 0.633 ± 0.082 & 0.818 ± 0.072 \\
            Augmented  &0.720 ± 0.061 & 0.646 ± 0.082 & 0.823 ± 0.067\\
            Clean &0.730 ± 0.058 & 0.663 ± 0.075 & 0.820 ± 0.072\\
			N4ITK	&0.721 ± 0.060 & 0.646 ± 0.078 & 0.823 ± 0.071\\
			NPP-Net	&0.710 ± 0.060 & 0.632 ± 0.080 & 0.822 ± 0.071\\
			CAE	    &0.721 ± 0.068 & 0.648 ± 0.089 & 0.823 ± 0.071\\
            ABCnet     &0.716 ± 0.061 & 0.640 ± 0.082 & 0.823 ± 0.071\\
			PHU-Net &0.721 ± 0.059 & 0.646 ± 0.078 & 0.825 ± 0.067\\
            VHU-Net &\textbf{0.724 ± 0.058} & \textbf{0.650 ± 0.077} & \textbf{0.825 ± 0.069}\\
			\bottomrule
		\end{tabular}
\end{table}


\subsubsection{Quantitative Comparison}
 As reported in Tables~\ref{Tab:transpro} and \ref{Tab:transpan}, bias field correction methods consistently yield notable improvements in segmentation metrics, including Dice, sensitivity, and PPV, relative to the original uncorrected images. On the prostate dataset, 
 with VHU-Net as the bias correction method, TransUNet achieves the highest Dice score of 0.756, representing a 10.20\% improvement over the uncorrected baseline (0.686). It also outperforms both classical (N4ITK: 0.697) and recent learning-based correction approaches (e.g., NPP-Net: 0.742, CAE: 0.702, PHU-Net: 0.694, ABCnet: 0.721). In terms of sensitivity and PPV, the VHU-Net-corrected images yield superior segmentation results of 0.684 and 0.857, respectively.
Performance gain is primarily attributed to the effectiveness of bias correction in reducing intensity inhomogeneities. These inhomogeneities often obscure tissue boundaries and diminish the contrast between anatomical structures. By restoring intensity consistency across similar tissue regions, bias field correction enables the segmentation model to perform more accurate feature learning and boundary localization, thereby enhancing segmentation accuracy.

Similar trends are observed in the abdominal dataset.
TransUNet demonstrates outstanding segmentation performance on VHU-Net corrected images, achieving a PPV of 0.825—slightly higher than on clean images (0.820). Its corresponding Dice score (0.724) and sensitivity (0.650) are also the closest to the clean image baseline (0.730 and 0.663, respectively), surpassing the results obtained with other bias correction methods such as N4ITK, CAE, NPP-Net, ABCnet, and PHU-Net.
These findings underscore the critical role of accurate bias field estimation and correction in supporting robust and precise medical image segmentation.

VHU-Net is compared with an additional baseline: TransUNet trained with RandomBiasField augmentations. This baseline is implemented using the PyTorch built-in RandomBiasField function. During training, 30\% of the samples are randomly augmented with bias fields generated by this function, with the order parameter $r=4$, consistent with Section~\ref{sec: Synthesis of bias field}. Quantitative results are summarized in Tables~\ref{Tab:transpro} and~\ref{Tab:transpan}. This augmentation strategy improves segmentation performance relative to the non-augmented TransUNet baseline. Nevertheless, the proposed bias-correction-then-segmentation pipeline achieves consistently superior performance. On the prostate dataset, the Dice score increases from 0.692 (augmented baseline) to 0.756 (ours), representing a relative improvement of 9.2\%. On the pancreas dataset, Dice improves from 0.720 to 0.724, corresponding to a 0.6\% relative gain. Similar upward trends are observed in PPV (from 0.850 to 0.857 for prostate, +0.8\%) and sensitivity (from 0.603 to 0.684, +13.4\%). These quantitative gains confirm that while the augmentation strategy enhances the segmentation model’s robustness to intensity inhomogeneity, residual artifacts may still interfere with reliable feature extraction. In contrast, bias field correction directly removes low-frequency intensity bias, producing more uniform tissue contrast and clearer anatomical boundaries. This cleaner input enables more effective and consistent feature learning, ultimately leading to improved segmentation performance. This suggests that bias correction and augmentation are complementary strategies, and integrating both may further improve downstream generalization in future work.

\begin{table}[htbp]
	\centering
	\caption{nnUNetV2 Prostate Segmentation Results.}
	\label{Tab:nnpros}
		\begin{tabular}{lccc} 
			\toprule
{Metrics}&{Dice $\uparrow$} &{Sensitivity $\uparrow$} &{PPV $\uparrow$} \\             
			\midrule
            \midrule
			Input	&0.841 ± 0.049 & 0.830 ± 0.083 & 0.862 ± 0.076\\
			N4ITK	&0.841 ± 0.049 & 0.830 ± 0.083 & 0.862 ± 0.076\\
			NPP-Net	&0.848 ± 0.051 & 0.812 ± 0.077 & \textbf{0.895 ± 0.061}\\
			CAE	    &0.841 ± 0.054 & 0.807 ± 0.092 & 0.892 ± 0.077\\
            ABCnet  &0.845 ± 0.051 & 0.819 ± 0.081 & 0.883 ± 0.080\\
			PHU-Net &0.843 ± 0.049 & 0.828 ± 0.082 & 0.869 ± 0.076\\
                VHU-Net &\textbf{0.855 ± 0.049} & \textbf{0.830 ± 0.075} & 0.889 ± 0.072\\
			\bottomrule
		\end{tabular}
\end{table}

\begin{table}[htbp]
	\centering
	\caption{nnUNetV2 Pancreas Segmentation Results.}
	\label{Tab:nnpan}
		\begin{tabular}{lccc} 
			\toprule
{Metrics}&{Dice $\uparrow$} &{Sensitivity $\uparrow$} &{PPV $\uparrow$} \\             
			\midrule
            \midrule
			Input	&0.778 ± 0.116 & 0.751 ± 0.173 & 0.844 ± 0.071\\
            Clean	&0.784 ± 0.112 & 0.762 ± 0.170 & 0.844 ± 0.073\\
			N4ITK	&0.784 ± 0.113 & 0.766 ± 0.173 & 0.839 ± 0.072\\
			NPP-Net	&0.783 ± 0.113 & 0.760 ± 0.171 & 0.844 ± 0.074\\
			CAE	    &0.784 ± 0.117 & \textbf{0.768 ± 0.175} & 0.837 ± 0.069\\
            ABCnet  &0.785 ± 0.113 & 0.767 ± 0.172 & 0.838 ± 0.070\\
			PHU-Net &0.782 ± 0.114 & 0.759 ± 0.173 & 0.843 ± 0.071\\
                VHU-Net &\textbf{0.785 ± 0.110} & 0.761 ± 0.168 & \textbf{0.844 ± 0.071}\\
			\bottomrule
		\end{tabular}
\end{table}

The prostate segmentation performance of nnUNetV2 after applying different bias field correction methods is summarized in Table~\ref{Tab:nnpros}. 
The uncorrected input and N4ITK produce identical Dice, sensitivity, and PPV values, respectively, showing that the traditional correction approach fails to improve segmentation performance. 
Compared with other learning-based approaches, VHU-Net delivers consistently superior segmentation performance. It improves the Dice score by around 1\% compared to NPP-Net and ABCnet, and by approximately 1.5\% over CAE and PHU-Net. 
Although nnUNetV2 achieves slightly higher PPV when using NPP-Net- and CAE-corrected images, this comes at the cost of reduced sensitivity compared with segmentation on the uncorrected input, suggesting that part of the true prostate boundary is missed. In contrast, VHU-Net maintains stable sensitivity while enhancing PPV, demonstrating its robustness in improving segmentation accuracy.
The reason is that VHU-Net provides a more accurate and homogeneous bias correction, which enhances the reliability of intensity features used by nnUNetV2. This leads to more accurate segmentation boundaries and improved reproducibility across different scans, underscoring the importance of robust bias correction as a preprocessing step for prostate segmentation. Furthermore, as shown in Table~\ref{Tab:nnpan}, VHU-Net continues to deliver consistent improvements on the pancreas segmentation task. VHU-Net achieves the highest Dice and sensitivity values, demonstrating its ability to further enhance nnUNetV2 performance even on challenging anatomical regions. Compared with other learning-based corrections, VHU-Net provides more stable and reliable gains, highlighting its strong generalization across different organs and imaging conditions.

\subsubsection{Qualitative Comparison}
\begin{figure*}[t]
	\centering
		\includegraphics[scale=0.18]{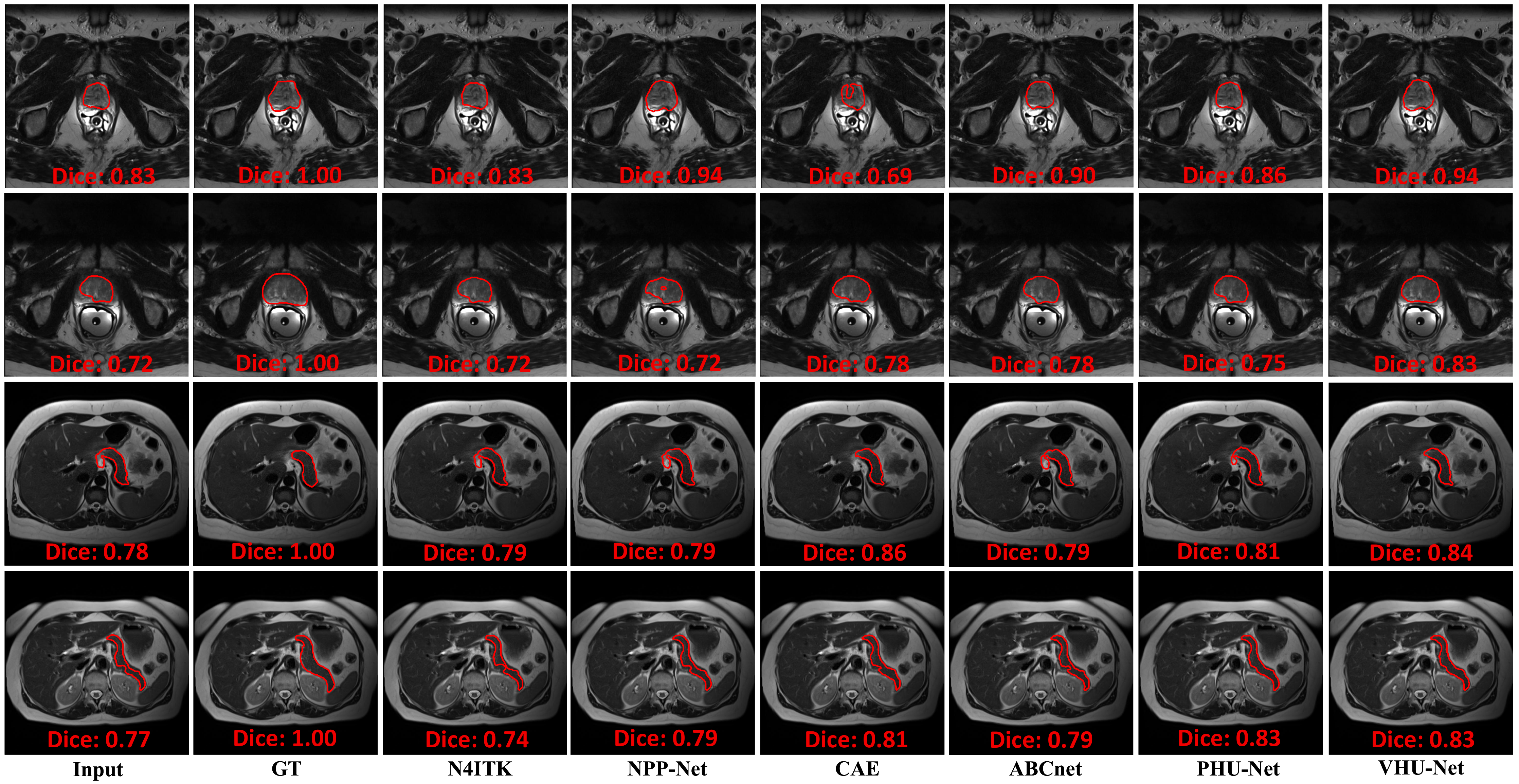}
	\caption{Qualitative results of segmentation experiments.}
	\label{Fig:visual_seg}
\end{figure*}

Fig.~\ref{Fig:visual_seg} illustrates the impact of various bias field correction methods on segmentation performance across both prostate and abdominal MRI images. From left to right, Fig.~\ref{Fig:visual_seg} presents the uncorrected input, the ground truth (GT) segmentation map, and the segmentation results after correction by N4ITK, NPP-Net, CAE, ABCnet, PHU-Net, and the proposed VHU-Net. Red contours indicate the predicted segmentations, overlaid on the original images. In the prostate examples, intensity inhomogeneity leads to suboptimal segmentations in the uncorrected input, where boundaries appear misaligned or incomplete. 
After applying N4ITK, CAE, ABCnet, or PHU-Net, the segmentation results still exhibit boundary leakage and inconsistent contours with noticeable deviations and irregular shapes. 
While NPP-Net enhances TransUNet segmentation performance in the first case, it fails to deliver consistent improvements across different scenarios.
In contrast, following correction by VHU-Net, TransUNet consistently achieves segmentations with accurate boundaries that closely match the ground truth, as evidenced by Dice coefficients of 0.94 and 0.83 on the respective test cases.
A similar trend is observed in the abdominal MRI images.
When corrected by competing methods, TransUNet presents noticeable distortions in the segmentation results, particularly along curved regions and pancreas boundaries. Conversely, VHU-Net enables more precise and faithful segmentations, with contours that tightly conform to the pancreas shape.
These results demonstrate that VHU-Net not only enhances intensity homogeneity but also facilitates improved downstream segmentation.

\begin{table}[t]
	\centering
	\caption{Analysis of Average Execution Time (in Seconds).}
	\label{Tab:time}
		\begin{tabular}{lcccc} 
			\toprule
{Dataset}&{PanSegData} &{HK} &{BIDMC} &{RUNMC}  \\ 
			\midrule
            \midrule
			N4ITK	&21.149   &173.569     &7.868    &84.966       \\
			NPP-Net	&0.937    &0.914      &0.970  & 0.552       \\
			CAE	    &0.231   &0.215      &0.301    & 0.163       \\
            ABCnet  &1.996        &1.863      &2.617    & 1.192         \\
			PHU-Net &0.506   &0.435     &0.669    &0.321        \\
                VHU-Net &0.987   &0.946      &0.973    &0.578        \\
			\bottomrule
		\end{tabular}
\end{table}

\subsection{Executive Time Analysis}
To assess the computational efficiency of different correction methods, we measured the average execution time required to process a single 3D MRI volume across four datasets. The results are summarized in Table~\ref{Tab:time}. Among all methods, the classical N4ITK method exhibits the highest runtime due to its iterative optimization scheme, with average times exceeding 20 seconds on PanSegData and surpassing 50 seconds on the RUNMC dataset. Such computational overhead limits its applicability in time-sensitive or large-scale settings. In contrast, deep learning–based methods offer substantially faster inference. 
Specifically, on the HK dataset, VHU-Net attains over 100-fold acceleration compared to N4ITK. Additionally, VHU-Net achieves near real-time performance, processing a single 3D volume from the RUNMC dataset in only 0.578 seconds. Besides, VHU-Net achieves a runtime comparable to that of NPP-Net. The reason is that the HT layer has a low computational complexity. The HT involves only additions and subtractions, resulting in zero multiply–accumulate operations (MACs). The subsequent scaling layer performs element-wise multiplication, resulting in $N^2$ MACs. In addition, multiplications involving the sign function in Eq.~(\ref{Eq: SST}) can be implemented using sign-bit operations. Therefore, the semi-soft thresholding layer requires $N^2$ MAC operations.
Consequently, the total MACs of the HT layer are on the order of $O(N^2)$, ensuring low computational cost.
Although CAE and PHU-Net exhibit lower runtime compared to VHU-Net, their bias field correction accuracy is also inferior, as shown in Tables~\ref{Tab:simupan},~\ref{Tab:realpana} and~\ref{Tab:realpros}. Therefore, VHU-Net achieves a trade-off between execution time and correction accuracy.



\section{Limitations and Future Work}
This study has several limitations that should be acknowledged. First, although the proposed method demonstrates strong performance in abdominal, breast, and prostate MRIs, its generalizability to other anatomies, such as the brain, heart, or knee, may be limited due to differences in bias field characteristics. The spatial distribution and magnitude of bias fields primarily depend on tissue electromagnetic properties, organ depth, and coil geometry. 
Therefore, the generalizability of the proposed method to other anatomies and imaging sequences remains to be systematically validated. Expanding the evaluation to larger multi-center cohorts with diverse acquisition protocols will be an important direction for future research.
Second, our model is subject to certain limitations. The performance depends on the choice of hyperparameters (e.g., $\varepsilon$, $\lambda$, $\delta$ ), which may require fine-tuning for different scanners or field strengths. The method may also underperform under extreme imaging conditions, such as severe motion or very low SNR. 
Third, the current evaluation of downstream tasks primarily relies on segmentation-based metrics to assess the effectiveness of the proposed bias field correction. To more comprehensively validate its impact, future studies should incorporate additional downstream tasks and clinical endpoints, such as classification performance and blinded radiologist reviews.

The present implementation is based on a 2D architecture. We adopt a 2D approach for two main reasons. All real datasets (from clinical studies) have variable slice thickness, requiring heavy resampling to uniform 3D grids, which greatly increases memory/runtime. Besides, even though the bias field is three-dimensional, its effect manifests in each 2D slice as a smoothly varying low-frequency artifact. By transforming the slice into the Hadamard domain, the proposed 2D model can effectively extract and suppress these low-frequency components, achieving accurate bias correction as shown in Table~\ref{Tab:simupan}. A 3D extension could leverage richer spatial features and potentially yield further performance gains. 
In the next logical step for validating and extending this framework, we consider cases like metal implants, which require dedicated artifact reduction
(MAR) techniques, to be a distinct challenge. 
Moreover, future work will focus on incorporating parameter optimization, domain adaptation, and foundation models to enhance cross-organ robustness.

\section{Conclusion}
This study presents VHU-Net, a novel frequency-aware variational architecture designed for effective bias field correction in body MRI. The proposed framework incorporates HT-based ConvHTBlocks in the encoder to separate low-frequency components associated with the bias field, while suppressing redundant high-frequency noise through a trainable scaling layer and a semi-soft thresholding mechanism. Moreover, the scaling layer applied in the HT domain can be mathematically interpreted as a dyadic convolution operation in the spatial domain. This connection allows VHU-Net to achieve the feature extraction capacity of a convolutional layer while maintaining the computational simplicity of element-wise operations.
The decoder integrates an IHTRTB to recover globally coherent and anatomically consistent bias fields by capturing inter-channel relationships. 
Multiple ConvHTBlocks also enhance the capacity of the decoder to reconstruct accurate bias fields. Additionally, a lightweight hypernetwork dynamically modulates feature responses at each decoding stage via channel-wise affine transformations, enhancing adaptability across anatomical variations. The variational formulation with an ELBO-based objective effectively regularizes the latent representations and enhances the accuracy of bias field estimation. 
The comprehensive evaluations across various body MRI datasets, including cross-center validation, demonstrate VHU-Net's robustness and clinical potential. By improving intensity uniformity, VHU-Net not only enhances visual image quality but also significantly improves downstream segmentation performance, potentially benefiting various clinical applications and quantitative analyses in medical imaging.

\appendix
\section{Derivation of a Differentiable Semi-Soft Thresholding Function}
\label{sec:dsst}
Semi-soft thresholding is a widely used technique in signal and image processing for attenuating low-amplitude coefficients while preserving those with significant magnitude in the frequency domain. The classical semi-soft thresholding function is defined as~\cite{lei2021remote}:
\begin{equation}
y = 
\begin{cases}
\text{sign}(x) \cdot \left( |x| - \dfrac{t}{e^{|x| - t}} \right), & |x| > t, \\
0, & |x| \le t,
\end{cases}
\label{eq:tmi_original}
\end{equation}
where \( \text{sign}(\cdot) \) denotes the sign function. $t$ represents the threshold parameter. $|\cdot|$ stands for the absolute value.

To derive a numerically stable and differentiable implementation suitable for neural network frameworks such as PyTorch, we decompose the function into two multiplicative components. We first define a binary gating function to suppress outputs below the threshold:
\begin{equation}
y_1 = \text{sign}\left( \text{ReLU}(|x| - t) \right),
\label{eq:y1}
\end{equation}
where $\text{ReLU}(\cdot)$ stands for the rectified linear unit (ReLU) function. It satisfies:
\begin{equation}
y_1 =
\begin{cases}
1, & |x| > t, \\
0, & |x| \le t.
\end{cases}
\end{equation}
Therefore, the binary gating function acts as a soft indicator function, enabling computation only when the input magnitude exceeds the threshold. 

After that, we define the activation value when the threshold is surpassed:
\begin{equation}
y_2 = \operatorname{sign}(x) \cdot \left( |x| - t \cdot e^{-(|x| - t)} \right).
\label{eq:y2}
\end{equation}
Combining Eq.~(\ref{eq:y1}) with Eq.~(\ref{eq:y2}), we derive a differentiable expression for the semi-soft thresholding function:
\begin{equation}
y = \text{sign}\left( \text{ReLU}(|x| - t) \right)\cdot\operatorname{sign}(x) \cdot \left( |x| - t \cdot e^{-(|x| - t)} \right).
\label{eq:yf}
\end{equation}
In PyTorch, both the sign and absolute value functions (torch.sign, torch.abs) are implemented with full support for automatic differentiation, ensuring that the overall formulation is differentiable and thus compatible with gradient-based optimization methods.





 \bibliographystyle{elsarticle-num} 
 \bibliography{cas-refs}

\end{document}